\documentclass[aps,prd,preprintnumbers,showpacs]{revtex4}
\usepackage{graphicx,epsfig}
\usepackage{subfigure}
\usepackage{amsmath}
\usepackage[usenames]{color}

\newcommand {\bc}{\begin {center}}
\newcommand {\ec}{\end {center}}
\newcommand {\be}{\begin {equation}}
\newcommand {\ee}{\end {equation}}
\newcommand {\beq}{\begin {eqnarray}}
\newcommand {\eeq}{\end {eqnarray}}
\newcommand {\unl}{\underline}
\newcommand {\ovl}{\overline}

\newcommand {\comment}[1]{}
\def\disp {\displaystyle}
\def\prt {\partial}
\renewcommand{\d}{{\rm d}}

\def\cL {{\cal L}}
\def\cM {{\cal M}}
\def\eps{\varepsilon}
\def\intl{\int\limits}
\def\f{\varphi}
\def\aa {{\rm a}}
\def\bb {{\rm b}}

\def\e {{\rm e}}
\def\ff {{f}}
\def\ii {{i}}

\def\x {{\rm x}}
\def\y {{\rm y}}
\def\z {{\rm z}}
\def\C {{\rm C}}
\def\R {{\rm R}}

\def\fs {{\rm fs}}

\def\ve {\textbf{\textit{e}}}
\def\vk {\textbf{\textit{k}}}

\def\vr {\textbf{\textit{r}}}
\def\vA {{\bf {A}}}
\def\vB {\textbf{\textit{B}}}

\def\valp{\boldsymbol{\alpha}}
\def\vgam{\boldsymbol{\gamma}}

\def\vomega{\boldsymbol{\omega}}
\def\vSig{\boldsymbol{\Sigma}}

\def\ue {\unl {e}}
\def\uk {\unl {k}}
\def\up {\unl {p}}
\def\ur {\unl {r}}
\def\uA {\unl {A}}
\def\uD {\unl {D}}

\def\ugam{\unl{\gamma}}
\def\unb{\unl{\nabla}}
\def\DR#1#2{{\frac {\d#1}{\d#2}}}
\def\Dr#1#2{\frac {\prt#1}{\prt#2}}

\def\lbr {\lambda\raise2pt\hbox {\hskip-4pt{\scriptsize --}}_\C}
\def\lbar {\lambda\hskip-5pt\raise3pt\hbox {--}} 

\begin{document}

\title{Compton scattering $S$-matrix and cross section in strong magnetic field}

\author{Alexander A. Mushtukov $^{1,2,3}$}
\email{al.mushtukov@gmail.com}
\author{Dmitrij I. Nagirner $^4$}
\email{dinagirner@gmail.com}
\author{Juri Poutanen $^{2,5}$}
\email{juri.poutanen@gmail.com}

\affiliation{
$^1$ Anton Pannekoek Institute, University of Amsterdam, Science Park 904, 1098 XH Amsterdam, The Netherlands \\
$^2$ Tuorla observatory, Department of Physics and Astronomy, University of Turku, V\"ais\"al\"antie 20, 21500 Piikki\"o, Finland \\
$^3$ Pulkovo Observatory of Russian Academy of Sciences, Saint-Petersburg 196140, Russia \\
$^4$ Sobolev Astronomical Institute, Saint Petersburg State University, Saint-Petersburg 198504, Russia \\
$^5$ Nordita, KTH Royal Institute of Technology and Stockholm University, Roslagstullsbacken 23, SE-10691 Stockholm, Sweden
}

\date{\today}

\begin{abstract}
Compton scattering of polarized radiation in a strong magnetic field is considered. The recipe for calculation of the scattering matrix elements, the differential and total cross sections based on quantum electrodynamic (QED) second order perturbation theory is presented for the case of arbitrary initial and final Landau level, electron momentum along the field and photon momentum. Photon polarization and electron spin state are taken into account. The correct dependence of natural Landau level width on the electron spin state is taken into account in general case of arbitrary initial photon momentum for the first time. A number of steps in calculations were simplified analytically making the presented recipe easy-to-use. The redistribution functions over the photon energy, momentum and polarization states are presented and discussed. The paper generalizes already known results and offers a basis for accurate calculation of radiation transfer in strong $B$-field, for example, in strongly magnetized neutron stars.
\end{abstract}


\maketitle
	
\section{Introduction}

Compton scattering is one of the most important processes of the interaction between radiation and matter in a number of astrophysical
objects. Strong external magnetic field significantly affects the properties of the scattering \citep{HL2006}: the interaction cross section becomes strongly dependent on energy, direction of photon momentum and polarization. It also depends on the magnetic field strength. A number of resonances corresponding to electron transition between the Landau levels appear. 
The resonant cross section value may exceed the Thomson scattering cross section $\sigma_{\rm T}$ by more than a factor of $10^6$.
All these factors have to be taken into account in the studies of radiation transfer and interaction between radiation and matter in strongly magnetized medium. Finally, Compton scattering plays a key role in formation of spectra from magnetized neutron star atmospheres 
\citep{1991ApJ...375L..49N,2003MNRAS.338..233H,2010ApJ...719..190W,2009A&A...500..891S,Poutanen2013,Putten2013} and 
dynamics of accretion onto magnetized neutron stars \citep{GS1973,MP1982,Musht2014,TsL2006,Staub2007}.

The simplest expressions for Compton scattering cross-section in strong $B$-field was derived in non-relativistic limit
by Canuto \cite{Canuto1971} and by Blandford \& Scharlemann \cite{BSh1976}. The non-relativistic treatment is limited to dipole radiation and therefore only scattering at the cyclotron fundamental is allowed. The non-relativistic approach works well when $k\gamma\ll m_{\rm e}c^2$, where $k$ is a photon energy, $m_{\rm e}$ and $\gamma$ are the electron rest mass and the Lorentz factor respectively. At higher energies the relativistic effects become important for calculations of the scattering cross section \citep{KN1929,T1930} and kinematics \citep{RL1979}.
The non-relativistic treatment is also limited to the magnetic field strength of $B\lesssim 10^{12}\,{\rm G}$ because the electron recoil becomes significant for higher $B$ \citep{DV1978}. 

The relativistic quantum electrodynamics (QED) treatment allows us to describe scattering at higher harmonics and also
consider the scattering which leads to electron transition to higher levels (so-called Raman scattering). It is the only way 
to describe scattering at high energies and strong magnetic field $B\gtrsim 10^{12}\,{\rm G}$, which is typical for young neutron stars.

The motion of electrons normal to the magnetic field is quantized in discrete Landau levels, whereas the longitudinal momentum can change continuously.  The particular case of Compton scattering with both initial and final electrons on the ground Landau level of zero initial velosity  was discussed by Herold \cite{Herold1979}. The scattering cross section from the ground to the arbitrary exited state was calculated  by Daugherty \& Harding \cite{DH1986} and by Meszaros \cite{M1992}. However, these QED calculations assume infinitely long-lived intermediate state and, therefore, are more relevant to photon energies far from the resonances. In order to calculate the resonant cross section one has to introduce a finite lifetime or decay width to the virtual electrons for cyclotronic transitions to lower Landau levels \citep{Pavlov1991}. For the specific case of ground-state to ground-state transition in the electron rest frame, when incident photons are parallel to the $B$-field, Gonthier et al. \citep{Gont2014} showed that the commonly used spin-average width of Landau levels does not correctly account for the spin dependence of the temporal decay and results in a wrong value of the cross section at the resonance as well as at very low photon energies, where the level width becomes comparable to the energy of the initial photon.

Scattering from the ground Landau level is commonly used as a basic approach in case of a strong field: 
$\hbar eB/(m_{\rm e}c)>k_{\rm B}T$, where $k_{\rm B}$ is the Boltzmann constant and $T$ is the electron temperature, when the majority of electrons occupy the ground energy level
\citep{PShM1989,1991ApJ...375L..49N,2007Ap&SS.308..109B,2008MNRAS.389..989N,2010ApJ...719..190W,2006hep.ph....9192C,Poutanen2013,Musht2014,2015arXiv150603600M}.
For the case of initial electron on the ground Landau level and the initial photon with momentum parallel to the magnetic field direction, the cross section has only one resonance and takes the simplest form. A simple approximation for the scattering cross section in this case was found by Gonthier et al. \cite{Gonth2000}. Their approximation represents the exact cross section quite well below the resonance and above it even for extremely strong fields ($B<10^{15}\,{\rm G}$).

Moving electrons scatter the photons differently because of relativistic effects. As a result, the electron distribution over momentum affects the exact cross section and broadens the resonance features. 
This effect could be important for formation of spectral features in X-ray pulsars \cite{2008ApJ...672.1127N,2011ApJ...730..106N} and for the estimations of radiation pressure \citep{GS1973,MP1982,Musht2014}, because the resonant scattering increases the effective interaction 
cross section dramatically. It is also important to use correct Landau level width and calculate correctly the exact resonant cross section here. 
The influence of electron distribution varies much with the photon momentum direction because electrons take part mostly in a motion along the $B$-field lines and the corresponding Doppler broadening varies a lot \cite{HD1991,Mu2012}. The scattering cross section for the case of thermal electrons was calculated and compared with cyclotron absorption by Harding \& Daugherty \cite{HD1991}. However, only polarization-averaged cross section for the case of initial electron at rest in the ground state was explored and an incorrect width of Landau levels based on Johnson-Lippmann wave-functions \citep{1949PhRv...76..828J} was taken into account (see \citep{Her1982} and \citep{Gont2014} for detailed discussion).

Description of additional effects such as vacuum polarization, two-photon scattering \citep{BLP1971}, pair creation \cite{2014Ap&SS.351..539W} demands the use of high order perturbation theory. They are beyond the scope of the present work.
However, it has to be pointed that the multiple photon scattering might be considered approximately as a chain of several elementary scatterings \cite{BMA1986}. Nevertheless, true scattering with an emission of two or more photons is a possibility which is given by QED treatment solely and the correct scattering cross section can be obtained only with relativistic treatment \citep{AM1991, SL2000}.

According to QED, the scattering process is described completely by its scattering matrix ($S$-matrix) \citep{BSh1976,BLP1971}, 
which contains the information about the probability amplitudes for the scattering. The transition probabilities and the effective cross sections of the various possible scattering are obtained from the $S$-matrix elements (which are complex numbers in general) as its squares, and therefore contain less information. The scattering cross sections are sufficient for a number of aims though, but the complete $S$-matrix is needed for general relativistic kinetic equation obtained recently by Mushtukov et al. \citep{Mu2012}.

In this paper we give a detailed scheme of calculation of Compton scattering $S$-matrix elements, the differential and the total cross-section based on the QED second order perturbation theory. Some steps were done analytically simplifying the calculations significantly and making them easy-to-use. The scheme is valid for arbitrary initial and final Landau level, though we focused on the scattering from the ground Landau level only. For the first time calculations do not assume restrictions on the photon momentum and electron distribution over momentum. As a result, the scheme could be applied to direct calculations of scattering by moving electrons, which is important for modeling of interaction between radiation and matter in the vicinity of accreting highly magnetized neutron stars \citep{1975A&A....42..311B,2007ApJ...654..435B,Musht2014,2015MNRAS.454.2714M}. 
The correct electron spin dependent Landau levels width \citep{Her1982,1986ApJ...309..372L,Pavlov1991} based on the Sokolov \& Ternov electron eigenfunctions of the magnetic 
Dirac equation \cite{ST1968,ST1986} for the first time is taken into account in a general case of arbitrary initial photon momentum. The correct spin dependent width was already used in calculations of Compton scattering cross section for the particular case of photons initially propagating along the magnetic field and ground-to-ground state transition of the electron \citep{Gont2014}. In our calculations we generalize this result. The correct Landau levels width is shown to be particularly important if we are interested in polarization of scattered photons and  accurate scattering cross section at the resonant energies \citep{Gont2014}. The obtained relations are valid in case of the magnetic field strength up to $\sim 10^{16}\,{\rm G}$ according to methods of particle description which are used in this paper (see Section \ref{sec:SPartDesc}).  We also discuss the redistribution function for the scattering (see Section \ref{sec:RedFun}), which traditionally are used in  radiation transfer equations and have a key role for studying the formation of spectral features near the cyclotron fundamental and its harmonics \citep{GDKK2011,Serber2000}. We provide a scheme of calculation of the cross section for the case of scattering by an ensemble of electrons described by any distribution function over momentum. The results could be used for the solution of the kinetic equation for Compton scattering obtained by Pavlov et al. \cite{PShM1989} and generalized by Mushtukov et al. \cite{Mu2012}. Since the general relativistic kinetic equation can be expressed via $S$-matrix elements only, we discuss some properties of scattering matrix elements which are important for the kinetic theory (see Section \ref{sec: S-matrix elements}). The paper describes the most general scheme for Compton scattering calculation in strong magnetic field based on the second order of QED perturbation theory and provides a ground for detailed investigation in a field of radiation transfer in case of strong external magnetic field.

We do not discuss here an influence of plasma effects on Compton scattering. The description of plasma effects was given in number of works \citep{BM1997,ChRSt2012}.

For simplicity we use the relativistic quantum system of units where the Planck constant, speed of light
and the electron mass are equal unity: $\hbar=c=m_{\rm e}=1$. In this case the length unit 
is Compton wavelength  $\lbar_\C=\hbar/m_{\rm e}c$, 
the unit of energy is the electron rest mass energy $m_{\rm e}c^2$, the frequency unit is
$m_{\rm e}c^2/\hbar$ and momentum is measured in $m_{\rm e}c$. The electron charge is $e=\sqrt{1/137.036}$.
The classical electron radius $r_\e$ is equal to the fine-structure constant $\alpha_\fs$  in using system of units: $r_\e=e^2/(m_{\rm e}c^2)=e^2/(\hbar c)=\alpha_\fs=1/137.036$.


\section{Particle description}
\label{sec:SPartDesc}

Let us consider constant and uniform magnetic field. The field is directed along the $z$-axis and could be represented by 3-dimensional vector
$\vB_\e=B_\e\,(0,0,1),$ where $B_\e>0$ is the field strength. 
Let us also use dimensionless magnetic field strength $b=B_\e/B_{\rm cr}$, which is a strength measured in units
of the Schwinger critical value $B_{\rm cr}=m^2_{\rm e}c^3/e\hbar=4.412\times 10^{13}$G.

\subsection{Electron in a strong magnetic field}

According to quantum mechanics the kinetic energy 
of the transverse motion is quantized in Landau levels \citep{LL1991}, since the particles gyrate in circular orbits.
Each electron is described by a set of quantum numbers which includes the Landau level number $n=0,1,2,...$, $z$-projection of electron 
momentum $p_{z}$, $y$-projection of electron momentum $p_{y}$ and electron spin projection onto the $z$-axis measured 
in $\hbar/2$-units $s=\pm 1$. We also use quantum number $\eps$ to describe the electron anti-particle - positron, $\eps=1$
for electrons and $\eps=-1$ for positrons. All Landau levels except the ground one ($n=0$) are degenerate with the spin-projection 
$s=\pm 1$. For the ground Landau level the spin degeneracy is one: $s=-1$.

The total electron energy in $B$-field with strength $b$ is defined by the Landau level number $n$ 
and $z$-projection of the electron momentum $p_{z}$:
\be \label{eq:RnZ}
E_n(Z)=\sqrt {1+p_{z}^2+2\,b\,n}.
\ee

According to relativistic quantum theory the electron states in external magnetic field are described by solutions of Dirac equation 
$\Psi_{n\,s}^\eps(\ur,p_{y},p_{z})$ enumerated by given quantum numbers (see Appendix \ref{sec:ElectronInMF}).
The solutions could be written in different ways. They could be found via the
eigenfunctions of a spin operator in the reference frame where the spin direction is fixed \citep{KM2003}.
In this case it is impossible to construct the Lorentz invariant amplitude for the processes with definite 
electron spin state since the spin direction is fixed. At the same time the amplitudes which are summed over the 
electron spin states are Lorentz invariant.
The solutions could be also found as the eigenfunctions of the operator 
$\hat{\mu}_z = m_{\rm e} \Sigma_z - \ii \gamma_0 \gamma_5 [\vSig \times \hat{p}]_z $, 
where  $\hat{p} =  - \ii {\bf \nabla} - e \uA_{\rm e}$ is the 
generalized momentum operator and $\uA_{\rm e} = B_{\rm e}(0, 0, x , 0)$ is 4-potential in Landau gauge 
\citep{ST1986} (see Appendix \ref{sec:DirMatr} for all necessary definitions). 
In this case the amplitudes for spin dependent processes are manifestly Lorentz-invariant \citep{KuzRum2013}.
Nevertheless, one could use both ways in case when we are interested only in the state averaged over 
the electron spin state. We discuss how to construct the electron wave function in Appendix \ref{sec:ElectronInMF}.

Further we will use following designations:
\be \label{eq:bnsn}
b_n=\sqrt {2\,b\,n},\,\,s_n=\sqrt {1+b_n^2}=\sqrt {1+2\,b\,n},\,\,
E_n(p_{z})=\sqrt {s_n^2+p_{z}^2}.
\ee

Let us choose that the laboratory reference frame as a frame where the initial electron has zero-velocity. 
Lorentz transformation along the magnetic field direction provides the conversion from one inertial system to another.

\subsection{Photon description}

Each photon is described by its energy $k$, the momentum direction defined by the unit-vector 
$\vomega=(\sin\theta\,\cos\f,\sin\theta\,\sin\f,\cos\theta)$ and its polarisation state. The 3-dimensional photon momentum 
in Cartesian coordinates: $\vk=(k_\x,k_\y,k_\z)=k\vomega$ and the corresponding photon 4-momentum: $\uk=\{k,\vk\}$.

The photon propagation in strong magnetic field is affected by vacuum polarization effects. Since photons may temporarily convert into virtual electron-positron pairs, which are polarized by the $B$-field, the dielectric and permeability tensors of magnetised vacuum are nontrivial. As a result the photon phase and group velocity depends on the polarization \citep{M1992,KM2003}, and it is natural to consider photons of two linear polarizations: $O$-mode (or $\|$-mode) photons which are linearly polarized in a plane containing $\vomega$ and $\vB$ and $X$-mode (or $\perp$-mode) photons which are polarized perpendicularly. 

The 4-vector potential for the photon can be defined as:
\be \label{eq:uAlur}
\uA_l(\ur)=\ue_le^{-i\uk\,\ur},\quad \ue_l=\{0,\ve_l\},\quad l=1,2.
\ee
The photon polarization is described in the co-ordinates which are specified by unit-vector $\vomega$ 
and two additional basis vectors:
$\ve_1=(\sin\f,-\cos\f,0)$ and $\ve_2=(\cos\theta\cos\f,\cos\theta\sin\f,-\sin\theta)$.
It is convenient to use so-called cyclic coordinates instead of Cartesian ones. The $z$-projection would be the same in this case, but
\be \label{eq:cyclcrd}
e_{1,\pm}=\sin\f\mp i\cos\f=\mp ie^{\pm i\f},\,\,
e_{2,\pm}=\cos\theta(\cos\f\pm i\sin\f)=\cos\theta e^{\pm i\f}
\ee
are used instead of $x$- and $y$-projections.
Thus, the coordinates of polarization basic vectors ($e_\z,\,e_{+},\,e_{-}$) in cyclic coordinates are 
\be \label{eq:ortscycl}
\ve_1(\vk)=(0,-ie^{i\f},ie^{-i\f}),\quad
\ve_2(\vk)=(-\sin\theta,\cos\theta e^{i\f},\cos\theta e^{-i\f}),\quad
\ve_3(\vk)=(\cos\theta,\sin\theta e^{i\f},\sin\theta e^{-i\f}).
\ee
The condition of orthonormality is $\ve_l(\vk)\ve_{l_1}^*(\vk)=\delta_{l\,l_1}$.

The photons are described here in the same manner as in case when the magnetic field is absent, i.e. we assume that the dispersion relation for the photons in magnetized vacuum does not differ from the dispersion relation for field absent case.
This approximation constrains the strength of the field. For estimations one 
needs to know vacuum dielectric tensor and the inverse permeability tensor for the case
of magnetized vacuum \citep{Adler1971,PLCh2004} but it is known that the indices of refraction differ from unity by more than $10\%$ only for
the fields hundred times stronger than the critical magnetic field, $b > 300$ \citep{Shaviv1999}. 
Thus, it restricts application of the developed formalism to $B\lesssim 10^{16}$G.

\section{Conservation laws and their consequences}
\label{sec:ConsLaws}

There are only three conservation laws for Compton scattering in strong $B$-field. It is the energy conservation law and the laws of longitudinal and transversal momentum conservation:
\be\label{eq:conslaws}
E_i+k_i=E_f+k_f,\,\,\,\,
p_{z,i}+k_i\cos\theta_i=p_{z,f}+k_f\cos\theta_f,\,\,\,\,
p_{y,i}+k_\ii\sin\theta_\ii\sin\varphi_\ii=p_{y,f}+k_\ff\sin\theta_\ff\sin\varphi_\ff,
\ee
where quantities which are corresponding to the initial particle states are denoted with the
index "$\ii$" while the quantities which are describing the final particle states are indexed with "$\ff$". 

In order to define the scattering event one has to define all quantum numbers which correspond to 
the particles in the initial and final states. The quantum numbers should comply with conservation 
laws (\ref{eq:conslaws}).
One possible way is to define all initial particle parameters and some final parameters.
The initial condition of a system could be defined by 
$p_{z,i},\,n_\ii,\,k_\ii$, two angles $\theta_\ii,\,\f_\ii$, and also by photon 
and electron polarization states. All other quantum number can be found from conservation laws (\ref{eq:conslaws}).
If one specifies the final Landau level $n_\ff$, then the final photon energy $k_\ff$ and the zenith angle $\theta_\ff$
comply with the following relation: 
$$
 k^2_\ff\sin^2\theta_\ff-2k_\ff(E_{\rm T}-Z_{\rm T}\cos\theta_\ff)+\left[E^2_{\rm T}-Z^2_{\rm T}-(1+2bn_\ff)\right]=0,
$$
where $E_{\rm T}\equiv E_\ii+k_\ii=E_\ff+k_\ff$ is the total energy and $Z_{\rm T}=Z_\ii+k_\ii\cos\theta_\ii=Z_\ff+k_\ff\cos\theta_\ff$ is
the total longitudinal momentum of electron-photon system.
In this case, the energy of the final photon is
\be \label{eq:kfres}
k_\ff=\frac{k_\ii\left[k_\ii\sin^2\theta_\ii+2(E_\ii-p_{z,i}\cos\theta_\ii)\right]+2b(n_\ii-n_\ff)}
{E_{\rm T}-Z_{\rm T}\cos\theta_\ff+(E_{\rm T}\cos\theta_\ff-Z_{\rm T})^2+s^2_\ff\sin\theta_\ff}.
\ee
Because the photon energy should be positive, there exist a limit on the final Landau level number: 
\be \label{eq:constrnf}
n_\ff\leq n_\ff^0=n_\ii+\lfloor k_\R/2b\rfloor,
\ee
where $\lfloor x\rfloor$ is a floor function of $x$ and $k_\R=k_\ii \left(k_\ii\sin^2\theta_\ii+2(E_\ii-p_{z,i}\cos\theta_\ii)\right)$.
Thus, the given initial photon ($k_\ii,\,\theta_\ii$) and electron ($p_{z,i},\,n_\ii$) parameters with the final 
Landau level number $n_\ff$ define uniquely the final photon energy in any direction.

\section{Matrix elements for Compton scattering}

\begin{figure}
\begin{minipage}{1.\linewidth}
\center{\includegraphics[angle=0, width=0.5\linewidth]{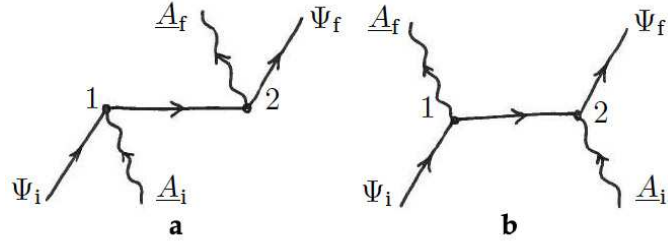} \\}
\end{minipage}
\caption{Two second order Feynman diagrams for Compton scattering. Wavy and streight lines depict photons and electrons/positrons
correspondingly. The streight line between two vertices corresponds to a virtual particle.}
\label{FeymanD}
\end{figure}

According to QED, Compton scattering is a second-order process and is described by two Feynman diagrams 
(Fig.\ref{FeymanD}). Both of them contain photon and electron before and after the interaction. The diagrams also contain 
so-called virtual electron/positron for which energy and momentum are not strictly conserved. 

Following the Feynman rules one gets elements of $S$-matrix for the process, which is the first step towards obtaining cross section \citep{BLP1971}. The initial electron is described by a particular solution of the Dirac equation for the electron in  external magnetic filed
$\Psi_\ii^{+}(\ur_1)=\Psi_{n_\ii s_\ii}^{+}(\ur_1,p_{y,i},p_{z,i})$ (see Section \ref{sec:Solution}). The final electron is also described by one of the solutions $\Psi_\ff^{+}(\ur_2)=\Psi_{n_\ff s_\ff}^{+}(\ur_2,p_{y,f},p_{z,f})$. The photons are described by 4-vectors of potential: 
$A_\ii(\ur),\,\,A_\ff(\ur)$ (see equation (\ref{eq:uAlur})). The interaction in a point $\ur$ 
gives us $e\ugam\uA(\ur)\Psi^{+}(\ur)$ according to the Feynman rules. The initial states give $e\ugam\uA_\ii(\ur)$ and
$\Psi^{+}_\ii(\ur)$, while the final ones give $e\ugam\uA_\ff^\dagger(\ur)$ and
$\Psi_\ff^{+\dagger}(\ur)\gamma^0=\ovl{\Psi}_\ff^{+}(\ur)$.

An internal electron line corresponds to the virtual electron state. The line begins in point $\ur_1$,
and ends in point $\ur_2$. The virtual particle is described by the relativistic propagator $G(\ur,\ur_1)$, which is a Green's function of the Dirac equation:
\be
\left[(i\unl{\bigtriangledown}+e\unl{A}_{\rm e})\ugam-1\right]G(\ur,\ur_1)=\delta(\ur-\ur_1).
\ee
It takes the following form for the case of electron in strong $B$-field:
\be \label{eq:prop}
G(\ur_2,\ur_1)=-{i}\sum_{n,\eps,s}\int\d p_{y}\,\d p_{z}
\Psi_{n s}^{\eps}(\vr_2,t_2,p_{y},p_{z})
\ovl{\Psi}_{n s}^{\,\eps}(\vr_1,t_1,p_{y},p_{z})\,\eps\,\Theta(\eps\,(t_2-t_1)),
\ee
where $\Theta(x)$ is the Heaviside step function. The same expression could be written in details as
\beq \label{eq:propag}
& \strut\disp G(\ur_2,\ur_1)=-\frac {i}{(2\pi)^2}
\intl_{-\infty}^\infty\d p_{z}e^{i\eps p_{z}(z_2-z_1)}\intl_{-\infty}^\infty
\d p_{y}e^{i\eps p_{y}(y_2-y_1)}\sum_{n=0}^\infty\sum_{s=\pm 1}
\sum_{\eps=\pm 1}\frac {1}{E_n(p_{z})} & \\
& \strut\disp \times e^{i\eps E_n(p_{z})(t_1-t_2)}
v_{n s}^\eps(\eps p_{z},x+\eps p_{y}/b)v_{n s}^{\eps\dagger}
(\eps p_{z},x+\eps p_{y}/b)\gamma^0\eps\Theta(\eps(t_2-t_1)), & \nonumber
\eeq
where the terms without any indexes correspond to the virtual electron/positron and one has to sum over the quantum numbers $n$ (Landau level) and $s$ (spin state).

Finally, the matrix element for Compton scattering which corresponds to the first two Feynman diagrams takes the following form:
\be \label{eq:Sfi}
S_{\ff\ii}=-4\pi ir_\e\int\d^4r_1\,\d^4r_2\ovl{\Psi}_\ff^{+}(\ur_2)\left\{
\left[\ugam\uA_\ff^\dagger(\ur_2)\right]\,G(\ur_2,\ur_1)\left[\,\ugam
\uA_\ii(\ur_1)\right]+\left[\,\ugam\uA_\ii(\ur_2)\right]\,G(\ur_2,\ur_1)\left[
\ugam\uA_\ff^\dagger(\ur_1)\right]\right\}\Psi_\ii^{+}(\ur_1).
\ee
The first term in curly brackets corresponds to the {\bf a}-diagram, while the second one corresponds to the {\bf b}-diagram (Fig.\ref{FeymanD}).

\section{Simplification and some algebra}
\label{sec:SimpAlg}

The general expression for the $S$-matrix element (\ref{eq:Sfi}) has to be specified. New expression for the elements will
contain integrals over time and space variables and sums over the discrete virtual electron/positron quantum numbers. 
It will be shown that 
the integrals could be calculated analytically as well as some sums. As a result we will get relatively simple expression for 
the elements of $S$-matrix, which would be suitable for the further analysis.

\subsection{First steps and integration over momentum and time variables}  

Using the expression for the electron propagator (\ref{eq:propag}) and the general expression for the
$S$-matrix elements (\ref{eq:Sfi}) we rewrite $S$-matrix elements in the following form:
\beq \label{eq:Sffii}
& \strut\disp S_{\ff\ii}\!=\!-4\pi r_\e\int\d^4r_1\,\d^4r_2\sum_{n,
\eps\,s}\int\d p_{y}\d p_{z}\Psi_{n_\ff\,s_\ff}^{+\dagger}(\ur_2,p_{y,f},p_{z,f})
\{[\gamma^0\ugam\uA_\ff^\dagger(\ur_2)]\,\Psi_{n\,s}^{\eps}
(\ur_2,p_{y},p_{z})\Psi_{n\,s}^{\eps\dagger}(\ur_1,p_{y},p_{z})
[\gamma^0\ugam\uA_\ii(\ur_1)] & \\
& \strut\disp +[\gamma^0\ugam\uA_\ii(\ur_2)]\,
\Psi_{n\,s}^{\eps}(\ur_2,p_{y},p_{z})\Psi_{n\,s}^{\eps\dagger}(\ur_1,p_{y},p_{z})
[\gamma^0\ugam\uA_\ff^\dagger(\ur_1)]\}\Psi^{+}_{n_\ii\,s_\ii}
(\ur_1,p_{y,i},p_{z,i})\eps\Theta(\eps(t_2-t_1)). & \nonumber
\eeq
We have also changed Dirac conjugated functions with the Hermit conjugated and $\gamma^0$ matrix has appeared as a result (see Appendix \ref{sec:DirMatr}). Using the expressions which are describing the electron (see Section \ref{sec:ElectronInMF}) and photon states we get:
\beq \label{eq:SSif}
& \strut\disp S_{\ff\ii}=-4\pi r_\e\sum_{n,\eps,s}\int\d p_{y}
\d p_{z}\int\d^4r_1\,\d^4r_2\frac {\eps\Theta(\eps(t_2-t_1))}{(2\pi)^2R_n}
\frac {1}{2\pi\sqrt {R_\ff}}v_{n_\ff\,s_\ff}^{+\dagger}(p_{z,f},x_2+p_{y,f}/b)
e^{i(E_\ff t_2-p_{y,f} y_2-p_{z,f} z_2)} & \nonumber \\
& \strut\disp \times\left\{
[\gamma^0\ugam\ue_{l_\ff}]e^{ik_\ff(t_2-\vomega_\ff\vr_2)}
v_{n\,s}^{\eps}(\eps p_{z},x_2+\eps p_{y}/b)v_{n\,s}^{\eps\dagger}
(\eps p_{z},x_1+\eps p_{y}/b)[\gamma^0\ugam\ue_{l_\ii}]
e^{-ik_\ii(t_1-\vomega_\ii\vr_1)} \right. & \nonumber \\
& \strut\disp \left. +[\gamma^0\ugam\ue_{l_\ii}]
e^{-ik_\ii(t_2-\vomega_\ii\vr_2)}v_{n\,s}^{\eps}(\eps Z,x_2+\eps p_{y}/b)
v_{n\,s}^{\eps\dagger}(\eps p_{z},x_1+\eps p_{y}/b)
[\gamma^0\ugam\ue_{l_\ff}]e^{ik_\ff(t_1-\vomega_\ff\vr_1)}\right\} & \\
& \strut\disp \times\frac {1}{2\pi\sqrt{E_\ii}}\,v_{n_\ii\,s_\ii}^{+}
(p_{z,i},x_1+p_{y,i}/b)e^{-i(E_\ii t_1-p_{y,i} y_1-p_{z,i} z_1)}
e^{\eps i[-E_n(t_2-t_1)+p_{y}(y_2-y_1)+p_{z}(z_2-z_1)]}. \nonumber &
\eeq

Taking the integrals over $y_1,y_2,z_1,z_2$ we get two products of $(2\pi)^4$ with four $\delta$-functions which are correspond to the momentum conservation in the Feynman diagram  vertices. For the first term ({\bf a}-diagram) we set
$\delta(\eps p_{y}-p_{y,f}-k_{\ff\y})\,\delta(p_{y,i}+k_{\ii\y}-\eps p_{y})\,
\delta(\eps p_{z}-p_{z,f}-k_{\ff\z})\,\delta(p_{z,i}+k_{\ii\z}-\eps p_{z})$, while for the second one ({\bf b}-diagram)
we have $\delta(\eps p_{y}-p_{y,f}+k_{\ii\y})\,\delta(p_{y,i}-k_{\ff\y}-\eps p_{y})\,
\delta(\eps p_{z}-p_{z,f}+k_{\ii\z})\,\delta(p_{z,i}-k_{\ff\z}-\eps p_{z})$. 
$(2\pi)^4$ has vanished. The integrals over $p_{y}$ and $p_{z}$ could be taken easily because of $\delta$-functions under the integrals. Finally, we are left with the product of two $\delta$-function $\delta(p_{y,i}+p_{y,f}-k_{\ii\y}-k_{\ff\y})
\delta(p_{z,i}+p_{z,f}-k_{\ii\z}-k_{\ff\z})$, which describe the 
conservation laws for the momentum. The values of $\eps p_{z}$ and
$\eps p_{y}$ for the virtual electron are different for two Feynman diagrams. Let us denote them with indexes 
"a" and "b" respectively:
\beq\label{eq:ZaZbYaYb}
& \strut\disp 
Z_\aa=p_{z,i}+k_{\ii\z}=p_{z,f}+k_{\ff\z},\,\,\,\,\,Y_\aa=p_{y,i}+k_{\ii\y}=p_{y,f}+k_{\ff\y},
 & \\
& \strut\disp 
Z_\bb=p_{z,i}-k_{\ff\z}=p_{z,f}-k_{\ii\z},\,\,\,\,\,Y_\bb=p_{y,i}-k_{\ff\y}=p_{y,f}-
k_{\ii\y}. \nonumber &
\eeq
The virtual electron energy would be also different for two diagrams: 
$E_{n\aa}=\sqrt{s_n^2+Z_\aa^2}$ and $E_{n\bb}=\sqrt{s_n^2+Z_\bb^2}$.

Thus, the expression for the matrix element (\ref{eq:SSif}) takes the form:
\beq \label{eq:SSiiff}
& \strut\disp S_{\ff\ii}=-4\pi r_\e\delta(p_{y,f}+k_{\ff\y}-
p_{y,i}-k_{\ii\y})\delta(p_{z,f}+k_{\ff\z}-p_{z,i}-k_{\ii\z})
\sum_{n,\eps,s}\int\d x_1\d x_2\d t_1\d t_2
\frac {\eps\Theta(\eps(t_2-t_1)}{\sqrt {E_\ii E_\ff}} & \nonumber \\
& \strut\disp \times v_{n_\ff\,s_\ff}^{+\dagger}(p_{z,f},x_{2\ff})
\left\{
e^{i[(E_\ff+k_\ff-\eps E_{n\aa})t_2+(\eps E_{n\aa}-k_\ii-E_\ii)t_1]}
M_{l_\ff}v_{n s}^\eps(Z_\aa,x_{2\aa})v_{n s}^{\eps\dagger}(Z_\aa,
x_{1\aa})M_{l_\ii}e^{i(k_{\ii\x}x_1-k_{\ff\x}x_2)} \right. \nonumber & \\ 
& \strut\disp \left. +e^{i(k_{\ii\x}x_2-k_{\ff\x}x_1)}
e^{i[(E_\ff-k_\ii-\eps E_{n\bb})t_2+(\eps E_{n\bb}+k_\ff-E_\ii)]}M_{l_\ii}
v_{n s}^\eps(Z_\bb,x_{2\bb})v_{n s}^{\eps\dagger}(Z_\bb,x_{1\bb})
M_{l_\ff}\right\}v_{n_\ii s_\ii}^{+}(p_{z,i},x_{1\ii}). &
\eeq
Here
\be \label{eq:Ml}
M_l=-\gamma^0\ugam\ue_l=\valp\ve_l=\left(\begin {array}{cccc}
0 & 0 & e_{lz} & e_{l-} \\ 0 & 0 & e_{l+} & -e_{lz} \\
e_{lz} & e_{l-} & 0 & 0 \\ e_{l+} & -e_{lz} & 0 & 0 \\
\end {array}\right),
\ee
where $l=1,2$ corresponds to photon polarization state and $\valp$ is given by equation (\ref{eq:valpSig}). 
In particular cases of $X$- and $O$-mode photons the matrices take the form:
$$ 
M_1=\left(\begin {array}{cccc}
0 & 0 & 0 & ie^{-i\f} \\
0 & 0 & -ie^{i\f} & 0 \\
0 & ie^{-i\f} & 0 & 0 \\
-ie^{i\f} & 0 & 0 & 0 \\
\end {array}\right),\quad
M_2=\left(\begin {array}{cccc}
0 & 0 & -\sin\theta & \cos\theta e^{-i\f} \\
0 & 0 & \cos\theta e^{i\f} & \sin\theta \\
-\sin\theta & \cos\theta e^{-i\f} & 0 & 0 \\
\cos\theta e^{i\f} & \sin\theta & 0 & 0 \\
\end {array}\right).
$$
In equations (\ref{eq:SSif},\ref{eq:SSiiff}) we also used notations for the spinor argument:
\beq \label{eq:argumspin}
& \strut\disp x_{2\,\ff}=x_2+Y_\ff/b,\quad x_{2\,\aa}=
x_2+(p_{y,f}+k_{\ff\y})/b,\quad x_{2\,\bb}=x_2+(p_{y,f}-k_{\ii\y})/b, & \\
& \strut\disp x_{1\,\ii}=x_1+p_{y,i}/b,\quad x_{1\,\aa}=
x_1+(p_{y,i}+k_{\ii\y})/b,\quad x_{1\,\bb}=x_1+(p_{y,i}-k_{\ff\y})/b. &
\eeq

The next step is taking the integrals over time variables $t_1$ and $t_2$. Using the relation $\int_{0}^{\infty} e^{ixy}\d y=\pi\delta(x)+{i}/{x}$
one gets the following expression for the case of {\bf a}-diagram:
\beq \label{eq:inta}
& \strut\disp \intl_{-\infty}^\infty\d t_1\intl_{-\infty}^\infty\d t_2
e^{i[(E_\ff+k_\ff-\eps E_{n\aa})t_2+(\eps E_{n\aa}-k_\ii-E_\ii)t_1]}
\eps\Theta(\eps(t_2-t_1)) & \nonumber \\
& \strut\disp\nonumber =2\pi\delta(E_\ff+k_\ff-E_\ii-k_\ii)\left[
\frac {i}{E_\ii+k_\ii-\eps E_{n\aa}}+\eps\delta(E_\ii+k_\ii-\eps E_{n\aa})
\right]=2\pi i\frac {\delta(E_\ff+k_\ff-E_\ii-k_\ii)}
{E_\ii+k_\ii-\eps E_{n\aa}}, &
\eeq
and for {\bf b}-diagram:
\beq \label{eq:intb}
& \strut\disp \intl_{-\infty}^\infty\d t_1\intl_{-\infty}^\infty\d t_2
e^{i[(E_\ff-k_\ii-\eps E_{n\bb})t_2+(\eps E_{n\bb}+k_\ff-E_\ii)t_1]}
\eps\Theta(\eps(t_2-t_1)) & \nonumber \\
& \strut\disp\nonumber =2\pi\delta(E_\ff+k_\ff-E_\ii-k_\ii)\left[\frac {i}
{E_\ii-k_\ff-\eps E_{n\bb}}+\eps\delta(E_\ii-k_\ff-\eps E_{n\bb})\right]=
2\pi i\frac {\delta(E_\ff+k_\ff-E_\ii-k_\ii)}{E_\ii-k_\ff-\eps E_{n\bb}}. &
\eeq

The $\delta$-functions outside the square brackets correspond to the energy conservation law. The $\delta$-function arguments inside the brackets (as well as denominators) describe the relation between the virtual particle energy and real 
particle energies in the Feynman diagrams vertexes. These arguments are not equal to zero in general, but may have values close to zero. 
It leads to the appearance of the resonances and matrix elements as well as cross sections become infinite. The infinities are removed by
the regularization procedure, when one takes into account the natural width of Landau levels \cite{Pavlov1991,Nag1993}
(see Section \ref{sec:ResonReg} and Appendix \ref{sec:LandauLevelWidth}). In this case the denominators are small 
but nevertheless differ from zero and the cross section values are not infinite.

After the integration over time variable the $S$-matrix elements (\ref{eq:SSiiff}) take the final form:
\beq \label{eq:Sxxif}
& \strut\disp S_{\ff\ii}=-8\pi^2ir_\e
\delta(p_{y,f}+k_{\ff y}-p_{y,i}-k_{\ii y})
\delta(p_{z,f}+k_{\ff z}-p_{z,i}-k_{\ii z})
\delta(E_\ff+k_\ff-E_\ii-k_\ii) & \\
& \strut\disp \times\sum_{n,\eps,s}\int\frac {\d x_1\d x_2}
{\sqrt {E_\ii E_\ff}}v_{n_\ff\, s_\ff}^{+\dagger}(p_{z,f},x_{2\ff})
\left\{M_{l_\ff}v_{n s}^\eps(Z_\aa,x_{2\aa})v_{n s}^{\eps\dagger}
(Z_\aa,x_{1\aa})M_{l_\ii}\frac {e^{i(k_{\ii x}x_1-k_{\ff x}x_2)}}
{E_\ii+k_\ii-\eps E_{n\aa}}\frac {1}{E_{n\aa}} \right. \nonumber & \\
& \strut\disp \left. +\frac {1}{E_{n\bb}}
\frac {e^{i(k_{\ii x}x_2-k_{\ff x}x_1)}}{E_\ii-k_\ff-\eps E_{n\bb}}M_{l_\ii}
v_{n s}^\eps(Z_\bb,x_{2\bb})v_{n s}^{\eps\dagger}(Z_\bb,x_{1\bb})
M_{l_\ff}\right\}v_{n_\ii s_\ii}^{+}(p_{z,i},x_{1\ii}), \nonumber &
\eeq
where the spinors $v^{\varepsilon}_{n s}$ are given by equations (\ref{eq:vnpp}-\ref{eq:vnmm}) (see Appendix \ref{sec:ElectronInMF}) and matrices $M_l$ are defined by equation (\ref{eq:Ml}). The braces in equation (\ref{eq:Sxxif}) contain $4\times 4$ matrices, while the whole construction under the integral is reduced to the complex function. The integration over $x_1$ and $x_2$ can be done analytically (see Section \ref{sec:IntegrationX}) as well as summation over the energy sign $\eps$ and spin state $s$ (see Section \ref{sec:SummES}). The summation over the electron spin states has to be done numerically.

\subsection{An integration over space variable in $S$-matrix elements} 
\label{sec:IntegrationX}

Let us take the expressions for spinors $v^{\varepsilon}_{n s}$ (\ref{eq:vnpp}-\ref{eq:vnmm})
(see Appendix \ref{sec:ElectronInMF}) and use them in final expression for the $S$-matrix elements (\ref{eq:Sxxif}).
Then the product of matrices $M_l$ and spinors $v^{\varepsilon}_{n s}$ under the integral in equation (\ref{eq:Sxxif}) is simplified and we are coming to the integrals which contain the products of $\chi$-function which are defined by equation (\ref{eq:chin}). 
All the integrals have the same form and could be represented via the Hermite polynomials:
\beq \label{eq:Il1l2x}
& \strut\disp I_{l_1\,l_2}=
\intl_{-\infty}^\infty\d x\chi^*_{l_1}(x+\alpha_1)
\chi^{\phantom{*}}_{l_2}(x+\alpha_2)e^{i\alpha x} & \nonumber \\
& \strut\disp =\frac{\sqrt {b}}{\sqrt {\pi}}\frac {i^{l_2-l_1}}
{\sqrt{2^{l_1}l_1!2^{l_2}l_2!}}\intl_{-\infty}^\infty\d x
e^{-b(x^2+2\alpha_1 x+\alpha_1^2+x^2+2\alpha_2 x+\alpha_2^2)/2+i\alpha x}
H_{l_1}(\sqrt {b}(x+\alpha_1))H_{l_2}(\sqrt {b}(x+\alpha_2)). &
\eeq
With new variables $u=\sqrt{b}[x+(\alpha_1+\alpha_2-i\alpha/b)/2]$ the last expression could be rewritten as
\beq \label{eq:Il1l2u}
I_{l_1\,l_2}=\frac {e^{\pi i(l_2-l_1)/2}}{\sqrt {\pi 2^{l_1+l_2}l_1!l_2!}}
\exp\left(-\frac {|\beta|^2}{4}-i\alpha\frac {\alpha_1+\alpha_2}{2}\right)
\intl_{-\infty}^\infty\d u\, e^{-u^2}H_{l_1}\left(u+\frac {\beta^*}{2}\right)
H_{l_2}\left(u-\frac {\beta}{2}\right),
\eeq
where $\beta=\sqrt {b}(\alpha_1-\alpha_2-i\alpha/b)$.
The integrals of the Hermitian polynomials product could be taken analytically and have well known expression through 
the Laguerre polynomials \citep{GR1980}:
$$
\intl_{-\infty}^\infty\d u\, e^{-u^2}H_n(u+x)H_m(u+y)=2^n\sqrt{\pi}m!x^{n-m}
L_m^{n-m}(-2xy),\,\,\,\,n\geq m.
$$
In our case $n=\max(l_1,l_2)=\ovl{l}$, $m=\min(l_1,l_2)=\unl{l}$, $n-m=
\ovl{l}-\unl{l}=|l_1-l_2|$, $xy=-|\beta|^2/4$, $x^{n-m}=
(|\beta|/2)^{|l_1-l_2|}e^{i(l_2-l_1)\arg\beta}e^{-\pi i(l_2-l_1+|l_1-l_2|)}$.
Therefore the integrals (\ref{eq:Il1l2u}) are transformed into simple expression: 
\be \label{eq:Il1l2res}
I_{l_1\,l_2}=\sqrt {\left(\frac {|\beta|^2}{2}\right)^{\ovl {l}-\unl {l}}}
\sqrt {\frac {\unl {l}!}{\ovl {l}!}}\exp\left(-\frac {|\beta|^2}{4}-i\alpha
\frac {\alpha_1+\alpha_2}{2}\right)\exp\left(i(l_2-l_1)\arg\beta-
\pi i\frac {\ovl {l}-\unl {l}}{2}\right)
L_{\unl {l}}^{\ovl {l}-\unl {l}}\left(\frac {|\beta|^2}{2}\right).
\ee

Let us separate real factors from the phase factors using following designations:
$$
\Xi_{l_1,l_2}(\beta)=\exp\left(i(l_2-l_1)\arg\beta-\pi i
\frac {\ovl{l}-\unl {l}}{2}\right),\quad \cL_{l_1,l_2}(\beta)=
\sqrt {\left(\frac {|\beta|^2}{2}\right)^{\ovl {l}-\unl {l}}}
\sqrt {\frac {\unl {l}!}{\ovl {l}!}}L_{\unl {l}}^{\ovl {l}-\unl {l}}
\left(\frac {|\beta|^2}{2}\right),
$$
so that
\be \label{eq:Il1l2}
I_{l_1\,l_2}=\exp\left(-\frac {|\beta|^2}{4}-i\alpha
\frac {\alpha_1+\alpha_2}{2}\right)\Lambda_{l_1,l_2}(\beta),\quad
\Lambda_{l_1,l_2}(\beta)=\Xi_{l_1,l_2}(\beta)\cL_{l_1,l_2}(\beta).
\ee

The expression for the matrix elements (\ref{eq:Sxxif}) contains four types of integrals over the space variables $x_1$ and $x_2$.
According to notation which is used in (\ref{eq:Il1l2x}) the integrals over $x_1$ contains the following parameters:
$$
x_{1,\aa},x_{1\ii}:\,\,\alpha_1=\frac {p_{y,i}+k_{\ii y}}{b},\,
\alpha_2=\frac {p_{y,i}}{b},\,\alpha=k_{\ii x};\quad\quad
x_{1\bb},x_{1\ii}:\,\,\alpha_1=\frac {p_{y,i}-k_{\ff y}}{b},\,
\alpha_2=\frac {p_{y,i}}{b},\,\alpha=-k_{\ff x},
$$
while in case of integrals over $x_2$ the parameters are:
$$
x_{2\ff},x_{2\aa}:\,\,\alpha_1=\frac {p_{y,f}}{b},\,
\alpha_2=\frac {p_{y,f}-k_{\ii y}}{b},\,\alpha=k_{\ii x};\quad\quad
x_{2,\ff},x_{2\bb}:\,\,\alpha_1=\frac {p_{y,f}}{b},\,
\alpha_2=\frac {p_{y,f}+k_{\ff y}}{b},\,\alpha=-k_{\ff x}.
$$
In the first combination $\beta=\beta_\ii$, while in the second case $\beta=\beta_\ff$, where 
\be \label{eq:betif}
\beta_\ii=-ik_\ii\sin\theta_\ii e^{i\f_\ii}/\sqrt {b},\quad
\beta_\ff=ik_\ff\sin\theta_\ff e^{i\f_\ff}/\sqrt {b}.
\ee 
The absolute value and the argument of $\beta_\ii$ and $\beta_\ff$ are
\be \label{eq:btmdari}
|\beta_\ii|=\frac {\sqrt {k_{\ii x}^2+k_{\ii y}^2}}{\sqrt{b}}=
\frac {k_\ii\sin\theta_\ii}{\sqrt{b}}=\frac {k_{\ii\perp}}{\sqrt{b}},\quad \arg\beta_\ii=
\f_\ii-\frac {\pi}{2},
\ee
\be \label{eq:btmdarf}
|\beta_\ff|=\frac {\sqrt {k_{\ff x}^2+k_{\ff y}^2}}{\sqrt{b}}=
\frac {k_\ff\sin\theta_\ff}{\sqrt{b}}=\frac {k_{\ff\perp}}{\sqrt{b}},\quad \arg\beta_\ff=
\f_\ff+\frac {\pi}{2}.
\ee

Finally, we get the following set of integrals in the expression for the $S$-matrix elements (\ref{eq:Sxxif}):
\beq \label{eq:intfa}
& \strut\disp \intl_{-\infty}^\infty\chi^*_{l_1}(x_{2\ff})
\chi^{\phantom {*}}_{l_2}(x_{2\aa})e^{-ik_{\ff x}x_2}\d x_2=
e^{-|\beta_\ff|^2/4+[ik_{\ff x}(p_{y,f}+k_{\ff y}/2)]/b}\Lambda_{l_1l_2}(\beta_\ff),
 & \\ \label{eq:intai} 
& \strut\disp \intl_{-\infty}^\infty\chi^*_{l_1}(x_{1\aa})
\chi^{\phantom{*}}_{l_2}(x_{1\ii})e^{ik_{\ii x}x_1}\d x_1=
e^{-|\beta_\ii|^2/4-[ik_{\ii x}(p_{y,i}+k_{\ii y}/2)]/b}\Lambda_{l_1l_2}(\beta_\ii),
 & \\ \label{eq:intfb}
& \strut\disp \intl_{-\infty}^\infty\chi^*_{l_1}(x_{2\ff})
\chi^{\phantom {*}}_{l_2}(x_{2\bb})e^{ik_{\ii x}x_2}\d x_2=
e^{-|\beta_\ii|^2/4-[ik_{\ii x}(p_{y,f}-k_{\ii y}/2)]/b}\Lambda_{l_1l_2}(\beta_\ii),
 & \\ \label{eq:intbi}
& \strut\disp \intl_{-\infty}^\infty\chi^*_{l_1}(x_{1\bb})
\chi^{\phantom {*}}_{l_2}(x_{1\ii})e^{-ik_{\ff x}x_1}\d x_1=
e^{-|\beta_\ff|^2/4+[ik_{\ff x}(p_{y,i}-k_{\ff y}/2)]/b}\Lambda_{l_1l_2}(\beta_\ff), &
\eeq
where the first couple of expressions corresponds to the {\bf a}-diagram, while the second couple to the {\bf b}-diagram.
Thus, the integration over the space variables is completed.

\subsection{Summation over the energy sign $\eps$ and spin state $s$ in the electron propagator} 
\label{sec:SummES}

The summation over the Landau levels has to be computed numerically, but the sums over the virtual particle energy sign and spin state are finite, and it is possible to find them analytically. Let us use an additional variable $V$, which is different for two Feynman diagrams: 
for the {\bf a}-diagram $V=V_\aa\equiv E_\ii+k_\ii$, and for the {\bf b}-diagram $V=V_\bb\equiv E_\ii-k_\ff$.
Then the term in the electron propagator corresponding to the $n$-th Landau level after summation over the energy sign and spin state takes the form (for both diagrams)
\be \label{eq:propagn}
\cM\equiv\frac {v_{n+}^{+}v_{n+}^{+\dagger}+v_{n-}^{+}v_{n-}^{+\dagger}}{V-E_n}
+\frac {v_{n+}^{-}v_{n+}^{-\dagger}+v_{n-}^{-}v_{n-}^{-\dagger}}{V+E_n}.
\ee
The sums in the nominators could be expressed using the commonly used matrices (see Appendix \ref{sec:DirMatr} 
for the designations). For the first term, we get
\beq \label{eq:proectel}
& \strut\disp v_{n+}^{+}(p_{z},x_2)v_{n+}^{+\dagger}(p_{z},x_1)+
v_{n-}^{+}(p_{z},x_2)v_{n-}^{+\dagger}(p_{z},x_1) & \nonumber \\
& \strut\disp =\frac{1}{2}\left[\chi_{n-1}(x_2)
\chi^*_{n-1}(x_1)(E_n\Sigma^{+}+p_{z}\alpha^{+}+D^{+})+
\chi_n(x_2)\chi^*_n(x_1)(E_n\Sigma^{-}-p_{z}\alpha^{-}+D^{-})\right.
 & \nonumber \\
& \strut\disp +\left.b_n\left(\chi_{n-1}(x_2)\chi^*_n(x_1)
\alpha_{+}+\chi_n(x_2)\chi^*_{n-1}(x_1)\alpha_{-}\right)\right],&
\eeq
and for the second term:
\beq
\label{eq:proectps}
& \strut\disp v_{n+}^{-}(p_{z},x_2)v_{n+}^{-\dagger}(p_{z},x_1)+
v_{n-}^{-}(p_{z},x_2)v_{n-}^{-\dagger}(p_{z},x_1) & \nonumber \\
& \strut\disp =\frac {1}{2}\left[\chi_{n-1}(x_2)
\chi^*_{n-1}(x_1)(E_n\Sigma^{+}+p_{z}\alpha^{+}-D^{+})+
\chi_n(x_2)\chi^*_n(x_1)(E_n\Sigma^{-}-p_{z}\alpha^{-}-D^{-})\right.
 & \nonumber \\
& \strut\disp -\left.b_n\left(\chi_{n-1}(x_2)\chi^*_n(x_1)
\alpha_{+}+\chi_n(x_2)\chi^*_{n-1}(x_1)\alpha_{-}\right)\right]. &
\eeq
At the same time, expression (\ref{eq:propagn}) could be reduced to
\be
\cM=\frac {1}{V^2-E_n^2}\left[V\sum_{s,\eps}v_{n,s}^\eps
v_{n,s}^{\eps\dagger}+E_n\sum_{s,\eps}\eps v_{n,s}^\eps
v_{n,s}^{\epsilon\dagger}\right],
\ee
where the sums in the square brackets contain only eight matrices:
\beq
& \strut\disp \sum_{s,\eps}v_{n,s}^\eps(p_{z},x_2)
v_{n,s}^{\eps\dagger}(p_{z},x_1)\!=\!\chi_{n-1}(x_2)\chi^*_{n-1}(x_1)
(E_n\Sigma^{+}\!+\!p_{z}\alpha^{+})\!+\!\chi_n(x_2)\chi^*_n(x_1)
(E_n\Sigma^{-}\!-\!p_{z}\alpha^{-}), & \\
& \strut\disp \sum_{s,\eps}\eps v_{n,s}^\eps(p_{z},x_2)
v_{n,s}^{\eps\dagger}(p_{z},x_1)=\chi_{n-1}(x_2)\chi^*_{n-1}(x_1)D^{+}+
\chi_n(x_2)\chi^*_n(x_1)D^{-} & \nonumber \\
& \strut\disp +b_n[\chi_{n-1}(x_2)\chi^*_n(x_1)\alpha_{+}+
\chi_n(x_2)\chi^*_{n-1}(x_1)\alpha_{-}]. &
\eeq
However, the obtained expressions are not always applicable, since it was assumed that $(V^2-E_n^2)\neq 0$, which is not generally satisfied. In case of $(V^2-E_n^2)=0$, which corresponds to resonant scattering, the situation is more complicated and is discussed separately (see Section \ref{sec:ResonReg}).

\section{Resonances: their position and regularization} 
\label{sec:ResonReg}

\begin{figure}[h]
\begin{minipage}[t]{0.49\linewidth}
\center{\includegraphics[angle=0, width=1.\linewidth]{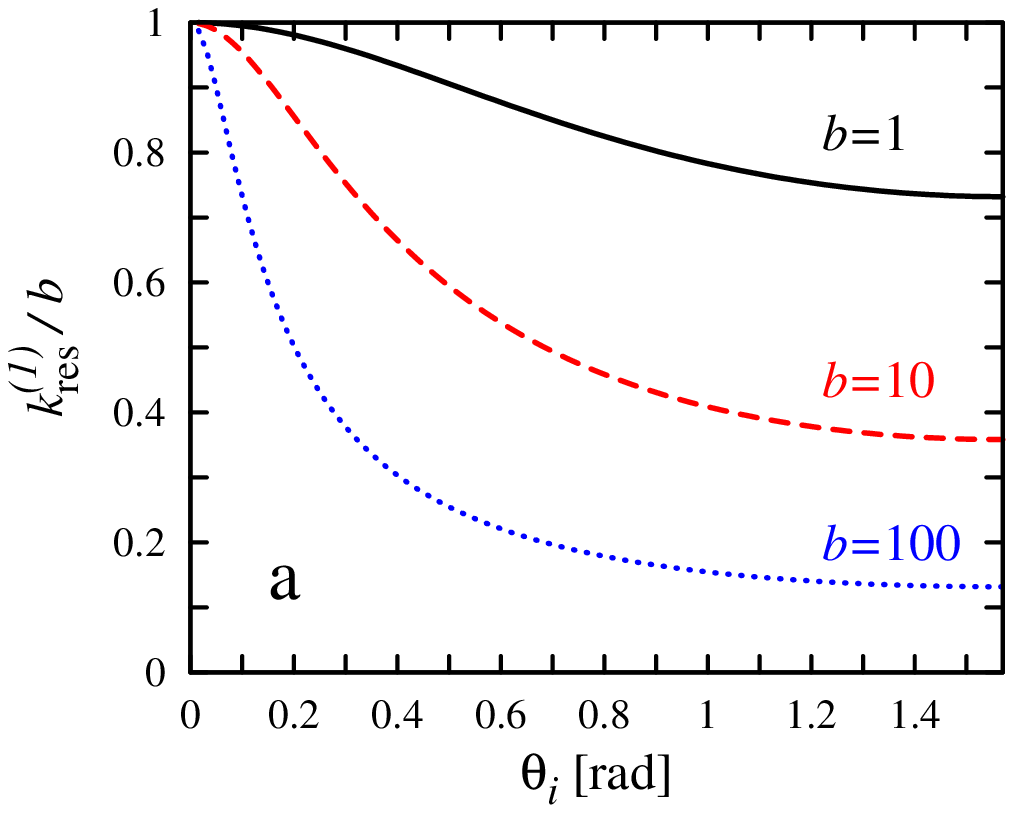} \\
}
\end{minipage}
\hfill
\begin{minipage}[t]{0.49\linewidth}
\center{\includegraphics[angle=0, width=1.\linewidth]{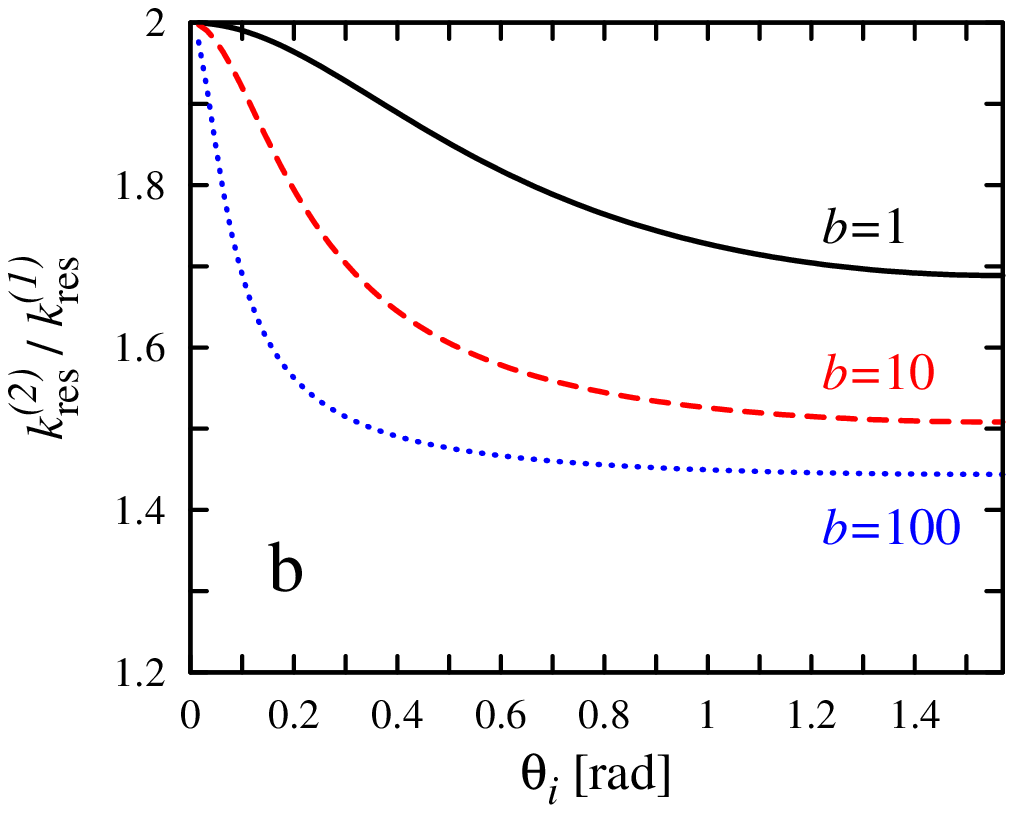} \\
}
\end{minipage}
\caption{Dependence of the resonance position on the direction. (a) Position of the first resonance as a function of initial photon momentum direction $\theta_i$ for different magnetic field strength $b$. The higher the field strength, the bigger the difference between the resonant energy in various directions.
(b) The ratio of the second and the first resonant energies as a function of photon momentum direction $\theta_i$ for various 
magnetic field strength $b$.}
\label{pic:reson}
\end{figure}

The differences $V_\aa-E_{n\aa}$ and/or $V_\bb-E_{n\bb}$ in (\ref{eq:propagn})
can become zeroes leading to the resonances in the cross sections. The resonance position
depends on the $B$-field strength, initial Landau level $n_{i}$, electron momentum along the field direction $p_{z,i}$
and the direction of the photon momentum:
\be
\label{eq:resonanceComp}
k^{(n)}_{\rm res}(b)=\frac{\sqrt{(E_\ii-p_{z,i}\cos\theta_\ii)^2+2b(n-n_\ii)\sin^2\theta_\ii}-(E_\ii-p_{z,i}\cos\theta_\ii)}
{\sin^2 \theta_\ii},\,\,\,\,\,\,\,
k^{(n)}_{\rm res}(b)>0.
\ee
If the electron occupies the ground Landau level and has zero-velocity along the $B$-field the expression 
for the resonance position simplifies:
\be 
\label{eq:resonanceSimp}
k^{(n)}_{\rm res}(b) \! =\!  
\left\{  \begin{array}{ll}
\strut\displaystyle \! \! 
\frac{\sqrt{1+2nb\sin^2\theta}-1}{\sin^2\theta}, & \mbox{for}\ \theta\neq0, \  n=1,2,..., \\
\! \!  b ,                                             & \mbox{for}\ \theta=0. 
 \end{array} \right. 
\ee
The resonance position depends on the photon momentum direction stronger in the case of stronger $B$-field (Fig. \ref{pic:reson}).
It is also obvious that the ratio of the resonant energies depends on the direction and the field strength (Fig. \ref{pic:reson}).

The resonances could be regularized if one takes into account the natural width of the Landau levels \citep{Pavlov1991,Nag1993}.
The width is defined by the electron transition rates from the occupied Landau levels and depends 
on the magnetic field strength, the Landau level number and 
the electron spin state \citep{Her1982,1986ApJ...309..372L,MZ1981,Pavlov1991,2005ApJ...630..430B} (see Appendix \ref{sec:LandauLevelWidth} for detailed discussion). The spin dependence of the Landau level width is particularly important 
if we investigate the polarization of scattered photons. Thus, there are two widths corresponding to each Landau level 
- $\Gamma_{n\,s}(p_{z})$.
The ground Landau level is an exceptional case. It has only one possible  spin state ($s=-1$) and its width 
$\Gamma_{0\,-}(p_{z})=0$, since the spontaneous transition $n_\ii=0\to n_\ff=0$ is impossible with any $p_{z}$. 
In order to regularise the resonances one should replace the energies of the initial and the final 
electrons $E_\ii$ and $E_\ff$ with $E_\ii-i\Gamma_\ii/2$ and
$E_\ff-i\Gamma_\ff/2$, the energy of virtual electron $E_n$
should be also replaced with $E_n+i\Gamma_{n\,s}/2$ \citep{Pavlov1991,Nag1993}. 

Let us define the following linear combination of the width of the Landau levels: 
\be \label{eq:Gamman}
\Gamma_n^\pm=\frac {\Gamma_{n\pm}}{2}+\frac {\Gamma_\ii+\Gamma_\ff}{4}.
\ee
Then the terms with the resonances (which we get from (\ref{eq:propagn})) 
in the propagator could be rewritten in the regularized form.
Let us also take into account the level width in the positron part.
Since the positron energy is $-E_n$ and the level width is positive, 
one should change $E_n$ by $E_n-i\Gamma_{n\,s}/2$ \citep{1993ApJ...412..351G}. Then the expression 
(\ref{eq:propagn}) can be rewritten as
\be \label{eq:regprop}\cM=
\frac{v_{n+}^{+}v_{n+}^{+\dagger}}{V-E_n-i\Gamma_n^{+}}+
\frac {v_{n-}^{+}v_{n-}^{+\dagger}}{V-E_n-i\Gamma_n^{-}}+
\frac {v_{n+}^{-}v_{n+}^{-\dagger}}{V+E_n-i\Gamma_n^{+}}+
\frac {v_{n-}^{-}v_{n-}^{-\dagger}}{V+E_n-i\Gamma_n^{-}}.
\ee
Useful relations for the spinor products in equation (\ref{eq:regprop}) are given in Appendix \ref{sec:ESpPr}.

Landau level natural width is also determines the scattering cross-section of photons with energy well below the cyclotron resonance, when $k_i\lesssim \Gamma_{n}$ \cite{Gont2014}. In this case the scattering cross section saturates at small constant value for the case of photons propagating along the magnetic field direction and for the case of photons of $X$-mode polarization of any angle between the $B$-field and photon momentum (see Appendix \ref{sec:LandauLevelWidth}).

\section{The S-matrix elements: phase factors}
\label{sec: S-matrix elements}

The elements of the scattering matrix are complex numbers in general and their phase factors are important in some cases: in particular it was shown by Mushtukov et al. \citep{Mu2012} that the exact form of the relativistic kinetic equation for polarized radiation demands the $S$-matrix elements and the cross section is not enough. Since we make a summation over the virtual electron Landau levels: $S_{\rm fi}=\sum_{n}S^{(n)}_{\rm fi}$, the phase factor depends on them. Nevertheless one can extract 
the phase factor $C_{\rm fi}$ ($S_{\rm fi}=C_{\rm fi}\sum_{n}S^{(n)}_{\rm fi}/C_{\rm fi}$) 
which does not depend on the variables describing the virtual electron 
(over which one make summation and integration in eq. (\ref{eq:Sxxif})):
\be \label{eq:common}
C_{\rm fi}=
\exp\left(i\frac{(\f_\ii+\f_\ff)(n_\ii-n_\ff)}{2}+ i\frac{(k_{\ff x}-k_{\ii x})(p_{y,i}+p_{y,f})}{2}\right).
\ee

The other phase factors are conjugated for {\bf a} and {\bf b} diagrams and depend on the virtual electron 
Landau level number over which the summation should be done. If we express the $S$-matrix elements with the relation 
$S_{\rm fi}=C_{\rm fi}\sum_{n}X^{(n)}_{\rm fi}M^{(n)}_{\rm fi}$, where $|X_{\rm fi}|=1$ and $M_{\rm fi}$ is a real number
($M_{\rm fi}\in\mathbf{R}$), then
\be \label{eq:exppm}
X^{(n)}_{\rm fi}=
\exp\left(\pm i\biggl(n-\frac {n_\ii+n_\ff}{2}\biggr)(\f_\ff-\f_\ii)\mp
i\frac {k_\ii k_\ff}{2}\sin\theta_\ii\sin\theta_\ff\sin(\f_\ff-\f_\ii)\right).
\ee
Upper and lower signs correspond to the {\bf a} and {\bf b} Feinmann  diagrams, respectively.
  
For the calculations of the matrix element one should know the following parameters:
the quantities which define the energy and momentum of initial particles - $n_\ii,\,p_{z,i}$
for the electron and $k_\ii,\,\theta_\ii$ for the photon, the quantities defining the energy and
momentum for the final particles - $n_\ff$ for the electron and $\theta_\ff,\,(\f_\ff-\f_\ff)$
for the photon. Some final quantities can be determined by the conservation laws (see Section \ref{sec:ConsLaws}).
The final Landau level should comply with the condition (\ref{eq:constrnf}). 
It is also necessary to specify the quantities which define the polarization state of the electrons 
in a final and initial states - $s_\ff,\,s_\ii$ and for the photon states
$l_\ff,\,l_\ii$. Then the recipe developed in Section \ref{sec:SimpAlg} allows us to transform expression
(\ref{eq:Sxxif}) and calculate the elements of the scattering matrix $S_{\rm fi}$.

The factors which are independent on the summation variable $n$ could be taken out from the 
summation sign. Their product is
\be
-\frac {8\pi^2ir_\e}{\sqrt {E_\ff E_\ii}}\delta(p_{y,f}+k_{\ff y}-p_{y,i}-
k_{\ii y})\delta(p_{z,f}+k_{\ff z}-p_{z,i}-k_{\ii z})\delta(E_\ff+k_\ff-E_\ii-
k_\ii)\frac {e^{-(|\beta_\ii|^2+|\beta_\ff|^2)/4}}
{4\sqrt {s_\ff s_\ii(E_\ff+s_\ff)(E_\ii+s_\ii)}}.
\ee

The obtained structure of $S$-matrix elements, which is given with (\ref{eq:common}) and (\ref{eq:exppm}), shows that 
the matrix elements are real numbers in case of scattering with only photon polarization change. It conforms to the structure
of general kinetic equation for Compton scattering in strong magnetic field obtained by Mushtukov et al. \citep{Mu2012}. 
The equation describes evolution 
of a density matrix kernel $\rho^{s'}_{s}(\vk,\vr,t)$ \citep{LLStat1980} and contains three items on the right hand side: 
$\uk\,\unl{\nabla}\rho^{s'}_{s}(\vk,\vr,t)=I_1+I_2+I_3$, where $\unl{\nabla}=(\partial/\partial t,-\nabla)$. 
The first two items describe photon redistribution over the polarization 
states only and the last term describes general photon redistribution over the energy, momentum and polarization states.
It was pointed that the first term contains the elements of $S$-matrix by themselves, while the second and the third therms contain 
usual products of matrix elements (as a result they could be rewritten using the cross sections, which is impossible for the
first term). Here we have shown that the matrix elements in the first term of the kinetic equation are real numbers and it would
simplify significantly the interpretation of physics behind this term.

\section{Cross sections and redistribution function} 

\subsection{Total and differential cross section}

As soon as one gets the $S$-matrix elements it is possible to find the scattering cross section. The differential Compton scattering 
cross section for the case of fixed initial electron state is
\beq
\label{eq:diffCS}
& \strut\disp
\frac{\d\sigma}{\d \Omega_\ff}(n_\ii,p_{z,i},s_\ii\,|\,k_\ii,\theta_\ii,l_\ii,\theta_\ff,l_\ff,\Delta\varphi_{\rm fi})=
& \nonumber \\
& \strut\disp
\sum\limits_{n_\ff, s_\ff}
\frac{E^2_\ii E^2_\ff}{(E_\ii+1)(E_\ff+1)}\frac{k_\ff}{k_\ii}
\frac{\sigma_{\rm T} |S_{\rm fi}|^2}{(E_\ii+k_\ii-k_\ff-\cos\theta_\ff(p_{z,i}+k_\ii\cos\theta_\ii-k_\ff\cos\theta_\ff))},
&
\eeq
where $\Delta\varphi_{\rm fi}=(\varphi_\ii-\varphi_\ff)$. Then the total cross section is obtained from the differential cross section
after the integration and summation over all possible final photon parameters:
\be
\label{eq:intCS}
\sigma_{l_\ii}(k_\ii,\theta_\ii; n_\ii,p_{z,i},s_\ii)
=\frac{1}{4\pi}
\sum\limits_{l_\ff}\int\limits_{0}^{\pi}\d\theta_\ff\,\sin\theta_\ff\int\limits_{0}^{2\pi}\d\varphi_\ff\frac{\d\sigma}{\d \Omega_\ff}.
\ee

\begin{figure}
\begin{minipage}{1.\linewidth}
\center{\includegraphics[angle=0, width=0.5\linewidth]{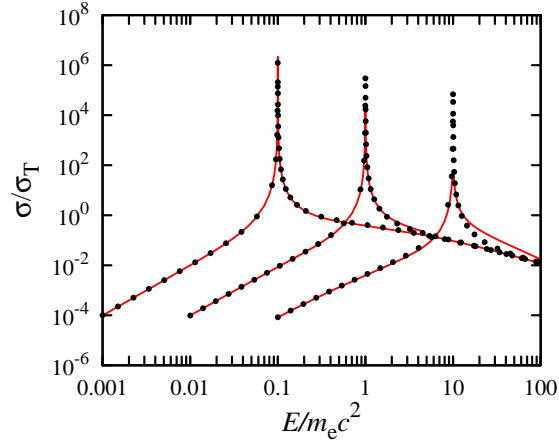} \\}
\end{minipage}
\caption{The exact cross-section for the photons which propagate initially along the $B$-field: $\theta_{\rm i}=0$ (red solid lines). 
There is no difference between polarizations in this case and only one resonance exists. Its position is defined by the field strength 
(\ref{eq:resonanceSimp}). Different curves correspond to various magnetic field strength: $b=0.1,\,1,\,10$ ($B\simeq 4.4\times 10^{12},\, \times 10^{13},\, \times 10^{14}$G). The scattering cross-section approximation obtained by Gonthier et al. \citep{Gonth2000} 
is given by black circles.
It works well, but overestimates the scattering cross-section near the resonant energy and underestimates the cross section after the
resonance in case of extremely high magnetic field strength: $b\gtrsim 10$.}
\label{pic_csB}
\end{figure}

\begin{figure}[h]
\begin{minipage}[t]{0.49\linewidth}
\center{\includegraphics[angle=0, width=1.\linewidth]{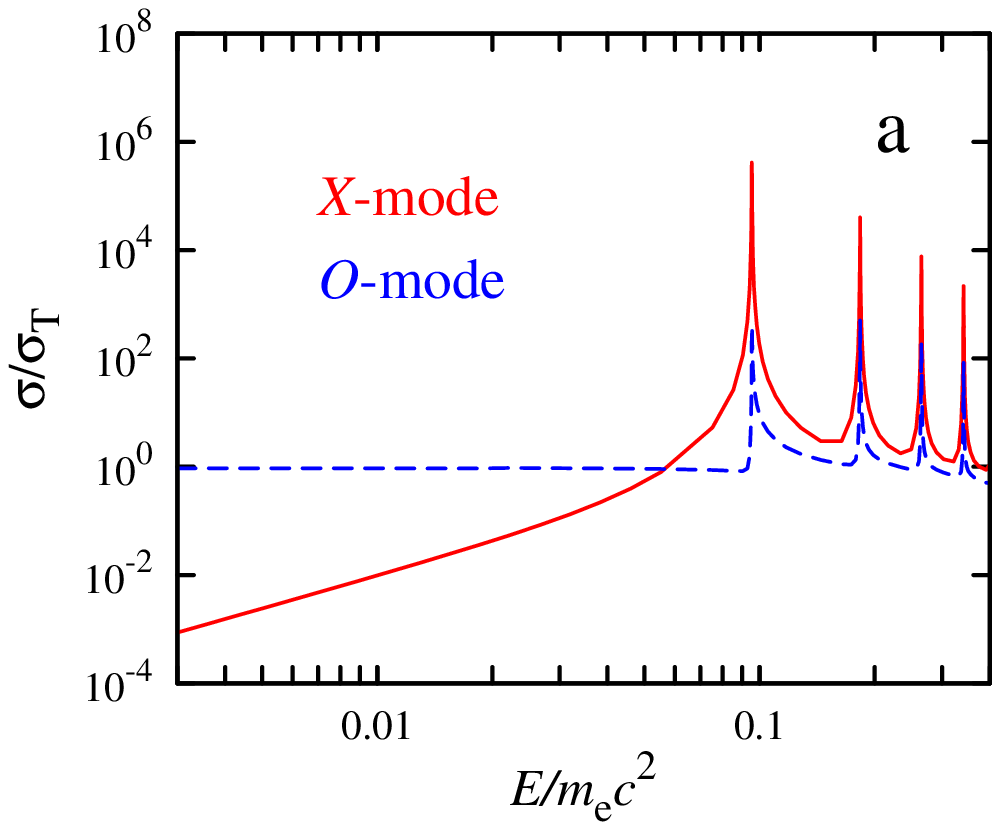}}
\end{minipage}
\hfill
\begin{minipage}[t]{0.49\linewidth}
\center{\includegraphics[angle=0, width=1.\linewidth]{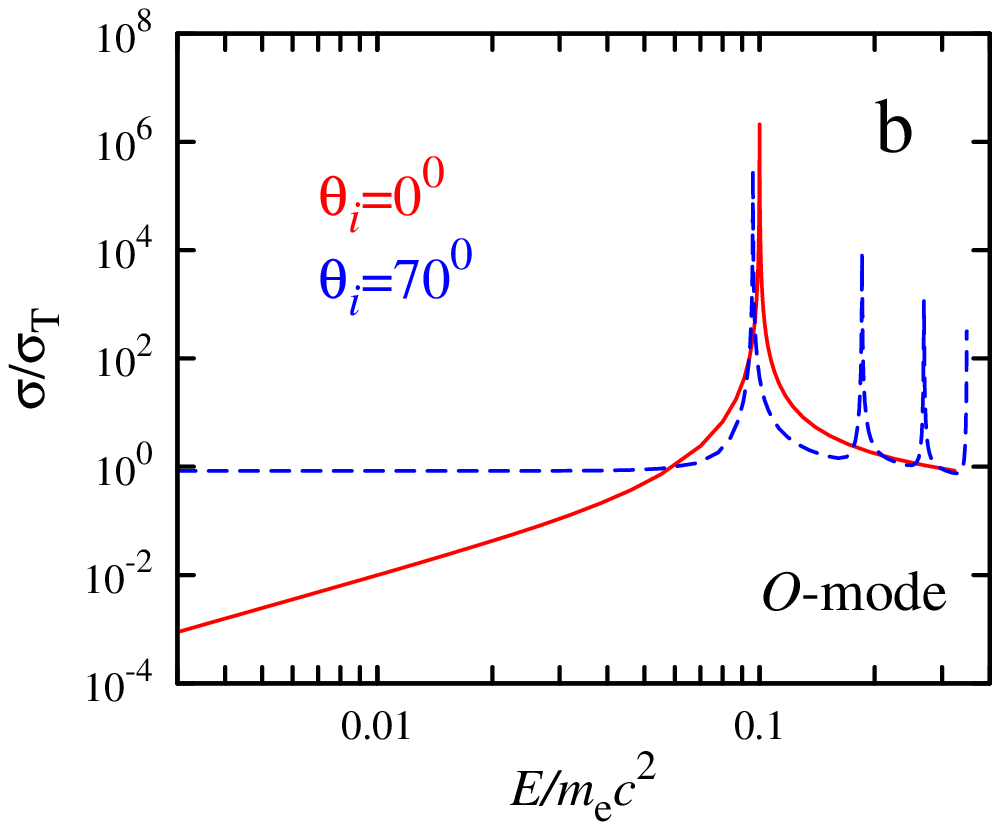}}
\end{minipage}
\caption{The cross section dependence on photon energy.
(a) The cross section for the $X$- and $O$-mode photons which propagate across the magnetic field ($\theta_\ii=\pi/2$) are given by solid red and dashed blue lines correspondingly. The cross section below the first resonance shows completely different behaviour. The resonance positions are almost the same, but the cross section of the resonant scattering is also different.
(b) The dependence of scattering cross section on the direction. For the case of $O$-mode photons of energies below the cyclotron energy $\sigma\propto (\sin^2\theta_\ii+(k/b)^2\cos^2\theta_\ii)$ if $b\lesssim 1$ and $k\ll b$.
Here $b=0.1$ ($B\simeq 4.4\times 10^{12}$ G).}
\label{pic_csAngT0}
\end{figure}

Examples of scattering cross section on the electron at rest are given in Fig.\ref{pic_csB} for the photon which propagates along the magnetic field and in Fig.\ref{pic_csAngT0} for the photons which propagate at some angle to the $B$-field. There is a number of resonances (\ref{eq:resonanceComp}) for the case of photons which propagate angularly to the magnetic field direction, while there is only one resonance for the case of photons which propagate along the field. The difference between $X$- and $O$-mode cross sections becomes stronger as the angle between the field direction and photon momentum increases.

In realistic situation, the electrons are distributed over the momentum, Landau level numbers and spin states.
In a case of sufficiently strong $B$-field ($k_{\rm B}T\ll E_{\rm cycl}$) one could assume that all electrons occupy 
the ground Landau level and therefore take part in one-dimensional motion and
have only one possible spin state ($s=-1$). In this case the differential cross section is defined  
by the electron distribution function $f_{n,s}(p_{z})$ (normalized as $\sum_{n,s}\int^{\infty}_{-\infty}\d p_{z} f_{n,s}(p_{z})=1$)
and the cross section corresponding to the scattering by an electron with given parameters (\ref{eq:diffCS}) is
\beq
\label{eq:diffCST}
& \strut\disp
\frac{\d\sigma^*}{\d \Omega_\ff}(k_\ii,\theta_\ii,l_\ii,\theta_\ff,l_\ff,\Delta\varphi_{\rm fi})=
\sum\limits_{n_\ii,s_\ii}\int\d p_{z,i}\,
\frac{\d\sigma}{\d \Omega_\ff}(n_\ii,p_{z,i},s_\ii\,|\,k_\ii,\theta_\ii,l_\ii,\theta_\ff,l_\ff,\Delta\varphi_{\rm fi})\,
f_{n_{\ii},s_{\ii}}(p_{z,i}).
&
\eeq
The total cross section in this case could be obtained from the differential one using relation (\ref{eq:intCS}).

\begin{figure}[h]
\begin{minipage}[h]{0.49\linewidth}
\center{\includegraphics[angle=0, width=1.\linewidth]{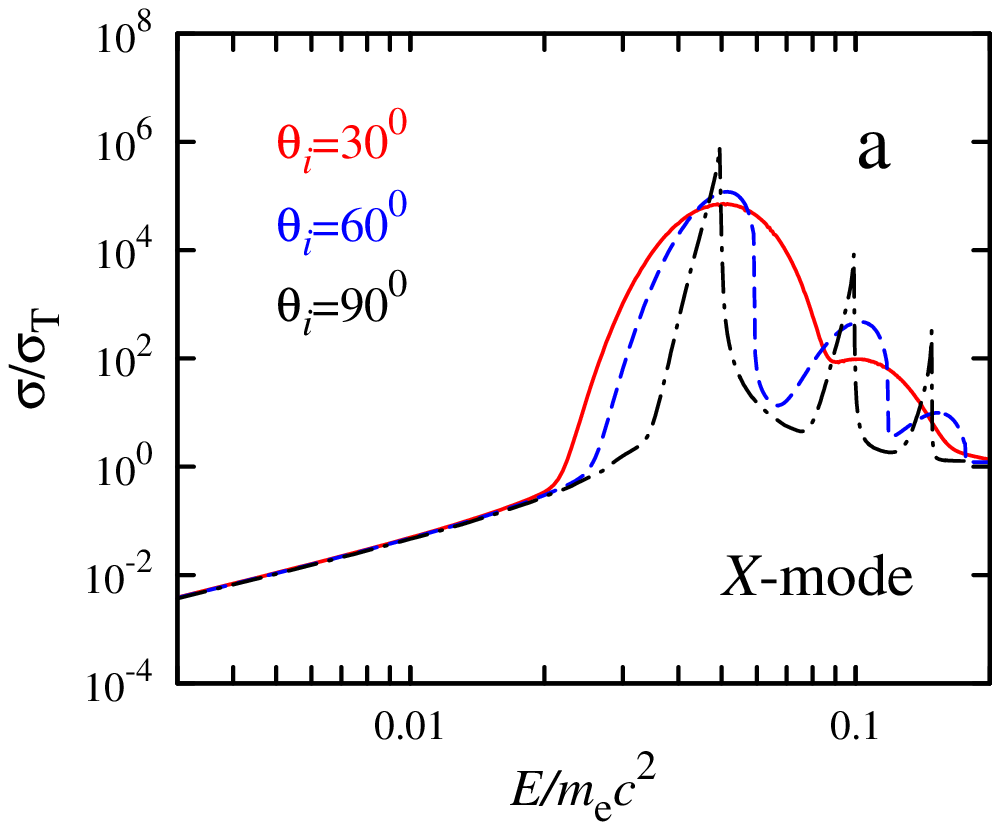}}
\end{minipage}
\hfill
\begin{minipage}[h]{0.49\linewidth}
\center{\includegraphics[angle=0, width=1.\linewidth]{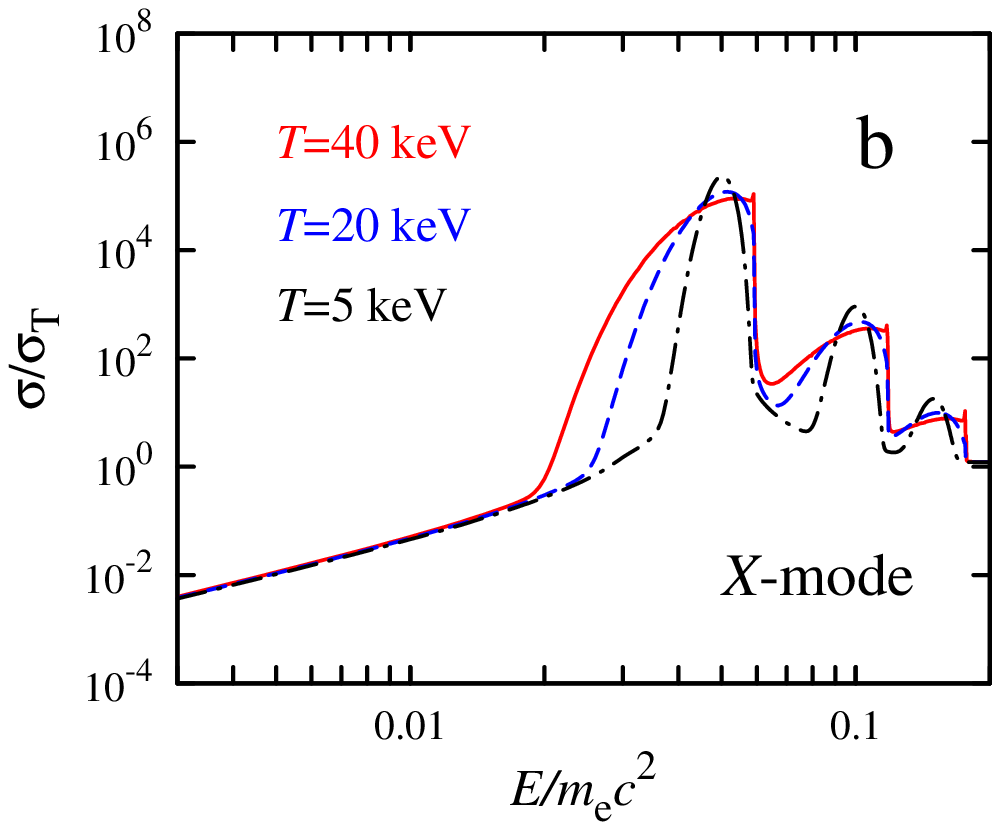}}
\end{minipage}
\caption{The cross section for the $X$-mode photons as a function of photon energy. (a) Dependence on incident angle $\theta_i$ for fixed electron temperature $T=20\,{\rm keV}$. (b) Dependence on electron temperature (for fixed $\theta_i=60^0$). The scattering features around the resonant energies are broadened with the width depending on the electron temperature and direction of photon momentum since the electrons mostly take part in one-dimensional 
motion. As a result the usual Doppler broadening takes place only for direction along the $B$-field, while in the perpendicular direction only 
the relativistic transversal Doppler broadening acts, and the scattering features are asymmetrical.
All results are given for $b=0.05$ ($B\simeq 2.2\times 10^{12}$G).}
\label{pic_csTermX}
\end{figure}

\begin{figure}[h]
\begin{minipage}[h]{0.49\linewidth}
\center{\includegraphics[angle=0, width=1.\linewidth]{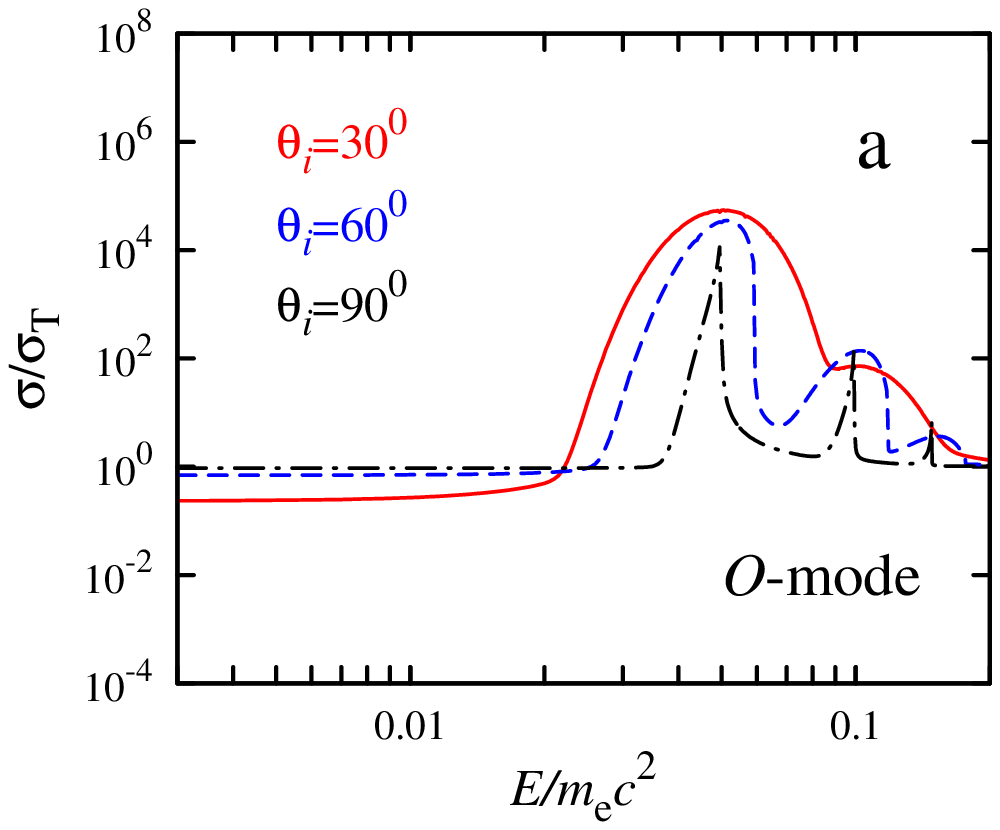}}
\end{minipage}
\hfill
\begin{minipage}[h]{0.49\linewidth}
\center{\includegraphics[angle=0, width=1.\linewidth]{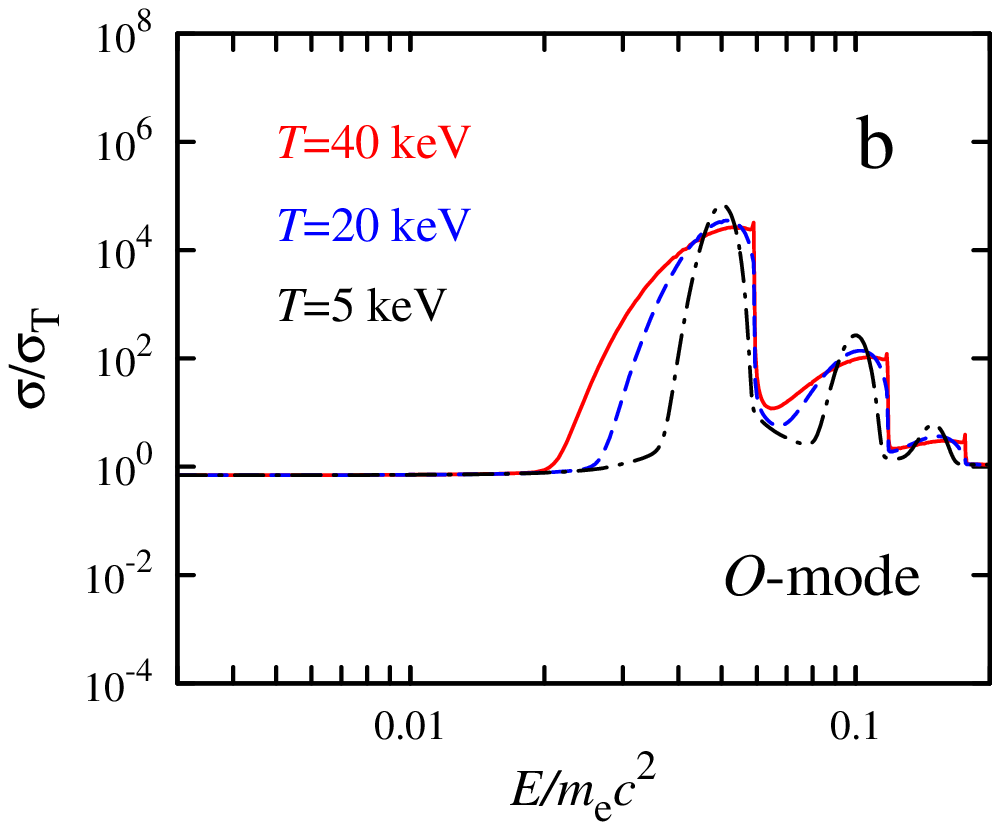}}
\end{minipage}
\caption{Same as Fig. \ref{pic_csTermX} but for the $O$-mode.}
\label{pic_csTermO}
\end{figure}

\begin{figure}[h]
\begin{minipage}[h]{0.33\linewidth}
\center{\includegraphics[angle=0, width=1.\linewidth]{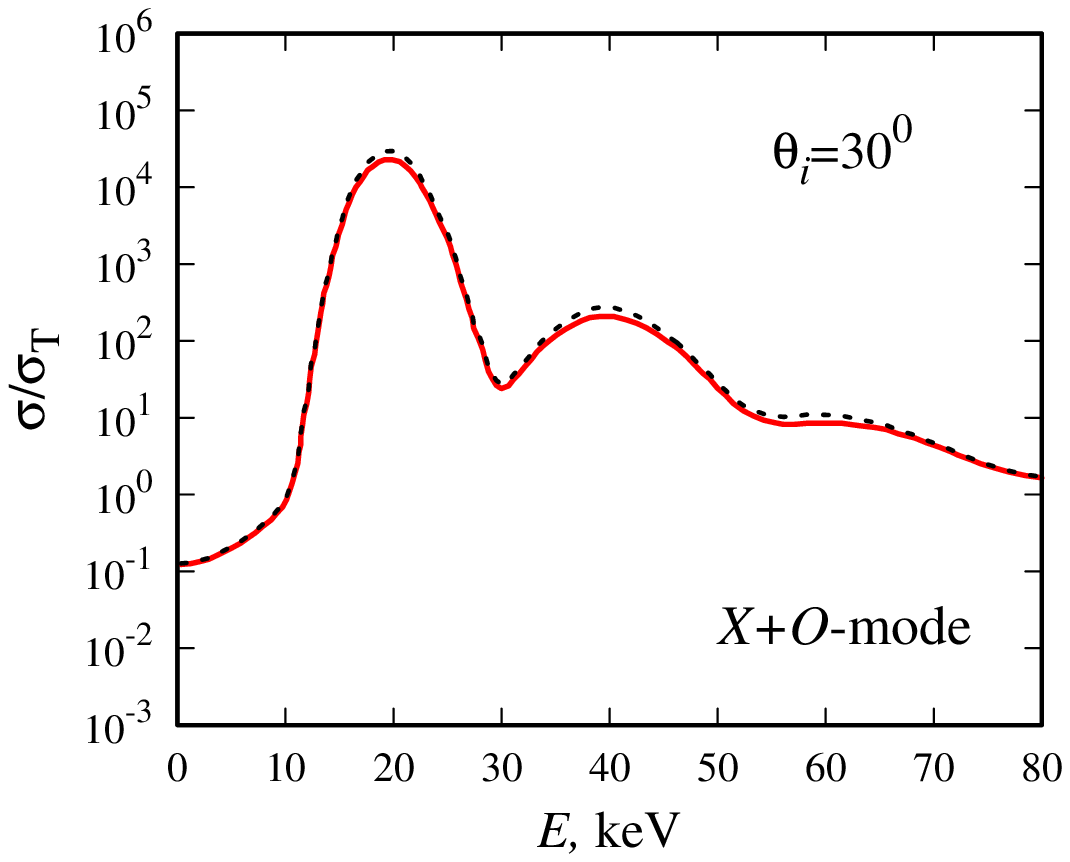}}
\end{minipage}
\hfill
\begin{minipage}[h]{0.33\linewidth}
\center{\includegraphics[angle=0, width=1.\linewidth]{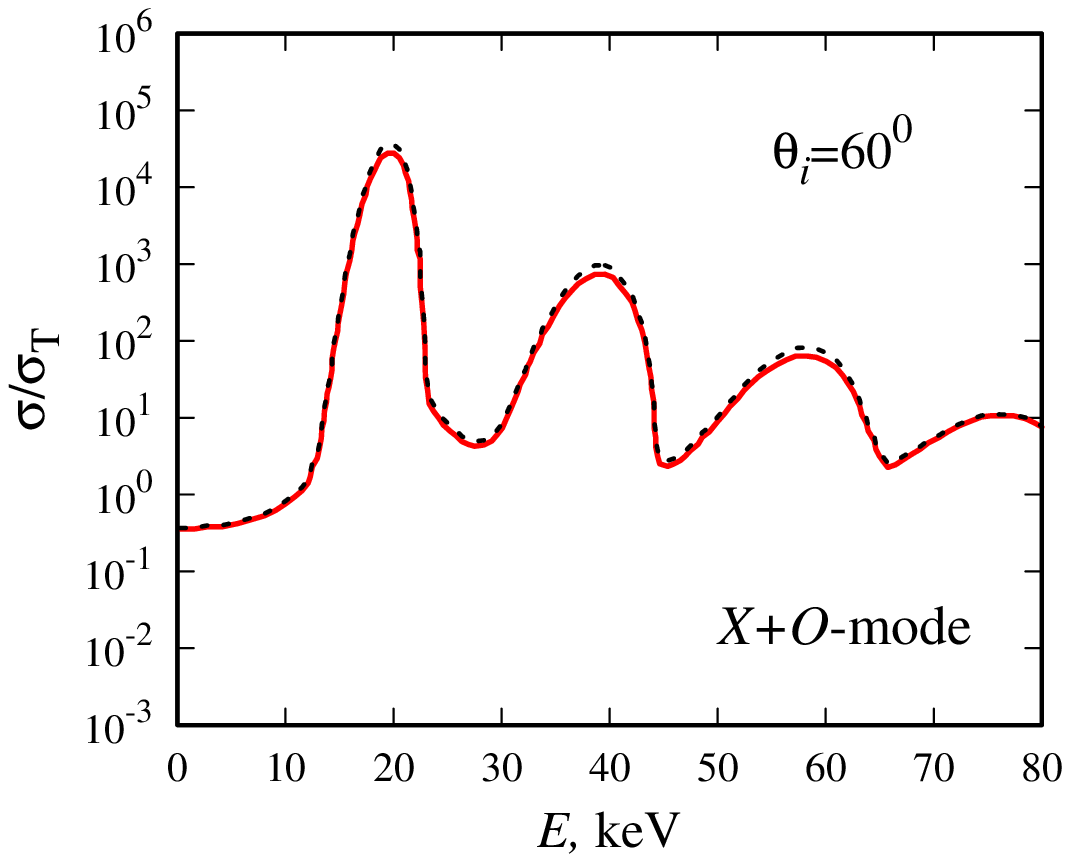}}
\end{minipage}
\hfill
\begin{minipage}[h]{0.32\linewidth}
\center{\includegraphics[angle=0, width=1.\linewidth]{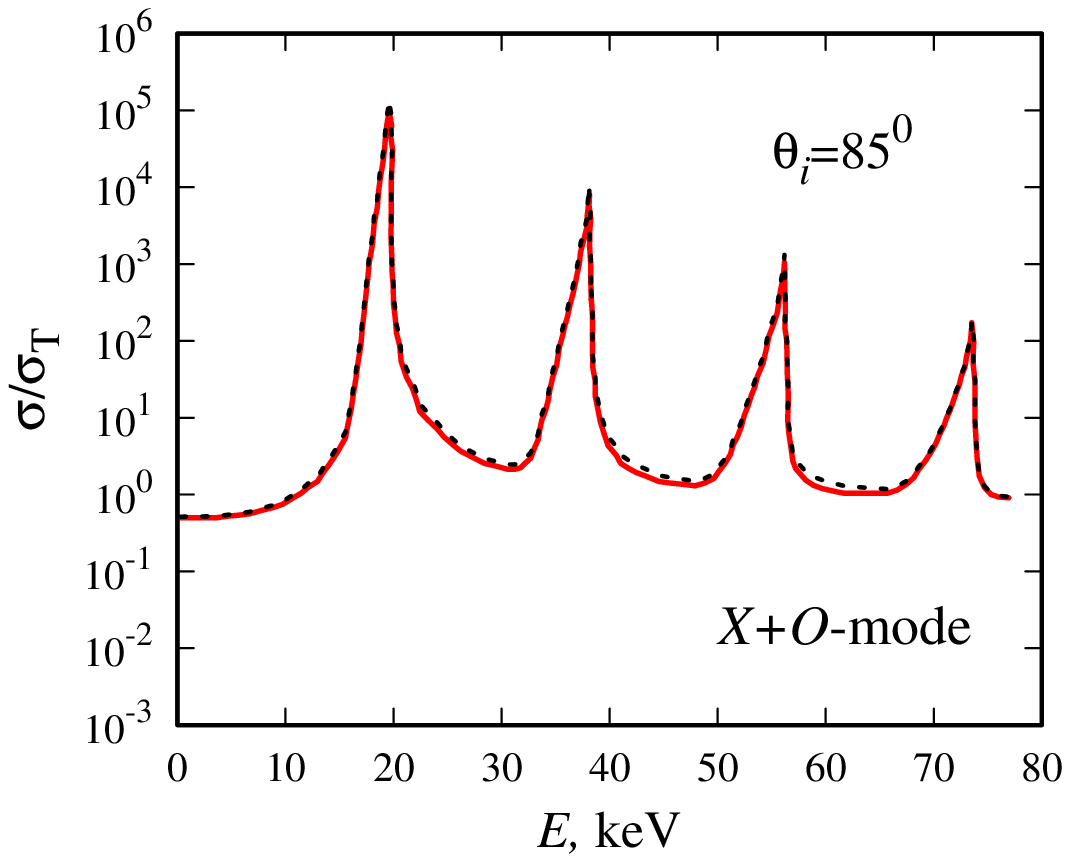}}
\end{minipage}
\caption{Polarization-averaged cross section for the case of magnetic field strength $B=1.7\times 10^{12}\,{\rm G}$, electron temperature $T=10\,{\rm keV}$ and various angles $\theta_i$ between the magnetic field direction and momentum of the initial photon is given by red solid lines. Black dashed line shows the results of the same calculations performed by Harding \& Daugherty \citep{HD1991}, where the Johnson-Lippmann wave functions \citep{1949PhRv...76..828J} were used.}
\label{pic_CompHD1991}
\end{figure}

Since the electrons in sufficiently strong magnetic field take part in one-dimensional motion, the
cross section near the resonant energies has special features. The shape of cyclotron features depends on the direction of initial photon momentum (see  Fig.\ref{pic_csTermX}(a) and \ref{pic_csTermO}(a)). For the case of longitudinal propagation, the ordinary Dopple broadening takes place. For other photons, the Doppler broadening is defined by the distribution of the projection of the electron momentum. The transversal 
Doppler effect becomes more important as the angle between the field direction and photon momentum increases. It provides asymmetrical
broadening of the cross section resonant features which is more evident for higher electron temperatures (see Fig.\ref{pic_csTermX}(b) and \ref{pic_csTermO}(b)). The results of our calculations are in agreement with the previously performed calculations \citep{HD1991} of scattering by thermal electrons (see Fig.\,\ref{pic_CompHD1991}). However, a small difference in the cross section at the resonance exists because the Sokolov-Ternov wave-function are used in our calculations instead of the Johnson-Lippmann wave functions \citep{1949PhRv...76..828J} (see \citep{Gont2014} for detailed discussion).

\begin{figure}
\begin{minipage}{1.\linewidth}
\center{\includegraphics[angle=0, width=0.5\linewidth]{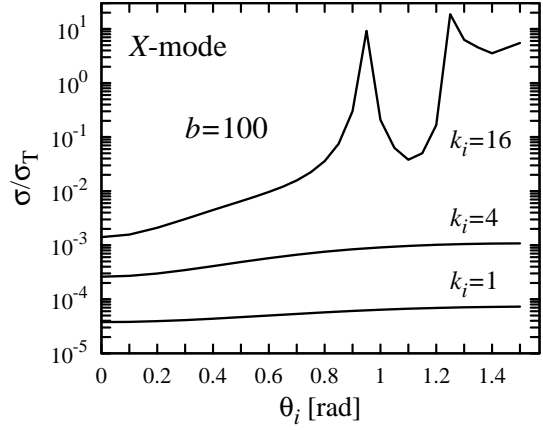} }
\end{minipage}
\caption{The scattering cross section as a function of initial angle between the photon momentum and magnetic field for a few initial photon energies. Since the resonant energies value depends strongly on the initial photon momentum in extremely high magnetic field,
the cross section shows a strong direction dependence as well even in case of the $X$-mode polarization. Here $b=100$ ($B\simeq 4.4\times 10^{15}$G).}
\label{pic:scSigAng}
\end{figure}

In extremely strong magnetic field ($b>10$ or $B\gtrsim 10^{15}$ G) some interesting features take place. The resonance position and
resonance energy ratios depend strongly on the direction (see Section \ref{sec:ResonReg}) and therefore the cross section 
depends strongly on the direction as well (Fig. \ref{pic:scSigAng}). It 
makes the problem of radiation transfer in magnetized plasma much more complicated. The dependence of resonance position on the field strength 
exist also for a relatively weak magnetic field, but it is not so dramatic. The resonant energies for the case of super-strong field 
are comparable or larger than the electron rest mass energy. As a result the decrease of relativistic cross section (``Klein-Nishina reduction") takes place.

\subsection{The redistribution function}
\label{sec:RedFun}

In order to use the results in astrophysical applications it is useful to construct the photon redistribution function 
describing the Compton scattering in strong $B$-field. The set of radiation transfer equations consists of two equations, one for 
each polarization mode ($l=1,2$):
\be
\frac{\d I_{l}(\vk)}{\d s}=
-\left(\alpha_{l}(\vk)+\kappa_{l}(\vk)\right)I_{l}(\vk)+\varepsilon_{l}(\vk)+
\sum\limits_{l'=1,2}\int\limits_{0}^{\infty}\d k'\int\limits_{0}^{\pi}\d\theta'\sin\theta'\int\limits_{0}^{2\pi}\d\varphi'
R(\vk',l'\longrightarrow \vk,l)I_{l'}(\vk'),
\ee
where $I_{l}(\vk)$ is an intensity in given polarization $l$ for the photons of momentum 
$\vk=k(\sin\theta\,\cos\f,\sin\theta\,\sin\f,\cos\theta)$, $\alpha_{l}(\vk)$ and $\kappa_l(\vk)$
are absorption coefficient due to the scattering process (or scattering coefficient) and true absorption correspondingly, 
$\varepsilon_{l}(\vk)$
is a true emission coefficient. The last item in the right hand side of the equation describes an emission due to the scattering 
processes in a given point and $R(\vk,l\longrightarrow \vk',l')$ is the redistribution function which defines 
the photon probability to change the 3-momentum and the polarization state in a scattering event. 
The redistribution function is normalized here in the following way:
\be
\sum\limits_{l'}\int\limits_{0}^{\infty}\d k'
\int\limits_{0}^{\pi}\d\theta'\sin\theta'
\int\limits_{0}^{2\pi}\d\varphi'
R(\vk,l\longrightarrow \vk',l')=
\alpha_l(\vk),
\ee
where the scattering coefficient $\alpha_l(\vk)=n_{\rm e}\sigma_{l}(k,\theta)$
and $n_{\rm e}$ is an electron concentration and $\sigma_{l}(k,\theta)$ is a scattering cross section.

According to the conservation laws, there is only one or several (for each admissible final Landau levels (\ref{eq:constrnf}))
possible final photon energies corresponding to each final photon direction in case of electron in a given quantum state, 
i.e. the final photon energy (\ref{eq:kfres}) 
is defined completely in case of fixed final scattering direction. The redistribution function over the zenith and azimuthal angles and polarization states is then
\be
\label{eq:RedFunAng}
R^{*}(k_\ii|\theta_\ii,\varphi_\ii,l_\ii\longrightarrow \theta_\ff,\varphi_\ff,l_\ff)
\equiv\int\limits_{0}^{\infty}\d k_\ff\, R(\vk_\ii,l_\ii\longrightarrow \vk_\ff,l_\ff)=
\frac{n_{\rm e}}{4\pi}\frac{\d\sigma}{\d \Omega_\ff},
\ee
where the differential cross section ${\d\sigma}/{\d \Omega_{f}}$ is given by equation (\ref{eq:diffCS}).

The general redistribution function, which corresponds to the scattering by the electron ensemble with a given distribution function over
the momentum, Landau levels numbers and spin states $f_{n_\ii,s_\ii}(p_{z,i})$ is:
\be
\label{eq:RedFunGen}
R(\vk_\ii,l_\ii\longrightarrow\vk_\ff,l_\ff)=
\frac{n_{\rm e}}{4\pi}\sum\limits_{n_\ii,s_\ii}\frac{\d\sigma}{\d \Omega_\ff}
(n_\ii,p_{z,i},s_\ii\,|\,k_\ii,\theta_\ii,l_\ii,\theta_\ff,l_\ff,\Delta\varphi_{\rm fi})
f_{n_\ii,s_\ii}(p_{z,i})\frac{\d p_{z,i}}{\d k_\ff},
\ee
where the $z$-projection of electron momentum is defined by the final photon energy: $p_{z,i}=p_{z,i}(k_\ff)$ and one could get it from the 
conservation laws (see Section \ref{sec:ConsLaws}). In case of scattering by electrons in a fixed state, the electron distribution 
function has to be replaced with $\delta$-function. It is easy to see that the 
integration over the final photon energy gives us the redistribution function over the directions only (\ref{eq:RedFunAng}) as it should.

\smallskip\smallskip\smallskip\smallskip

\begin{figure}
\begin{minipage}{0.49\linewidth}
\center{\includegraphics[angle=0, width=1.\linewidth]{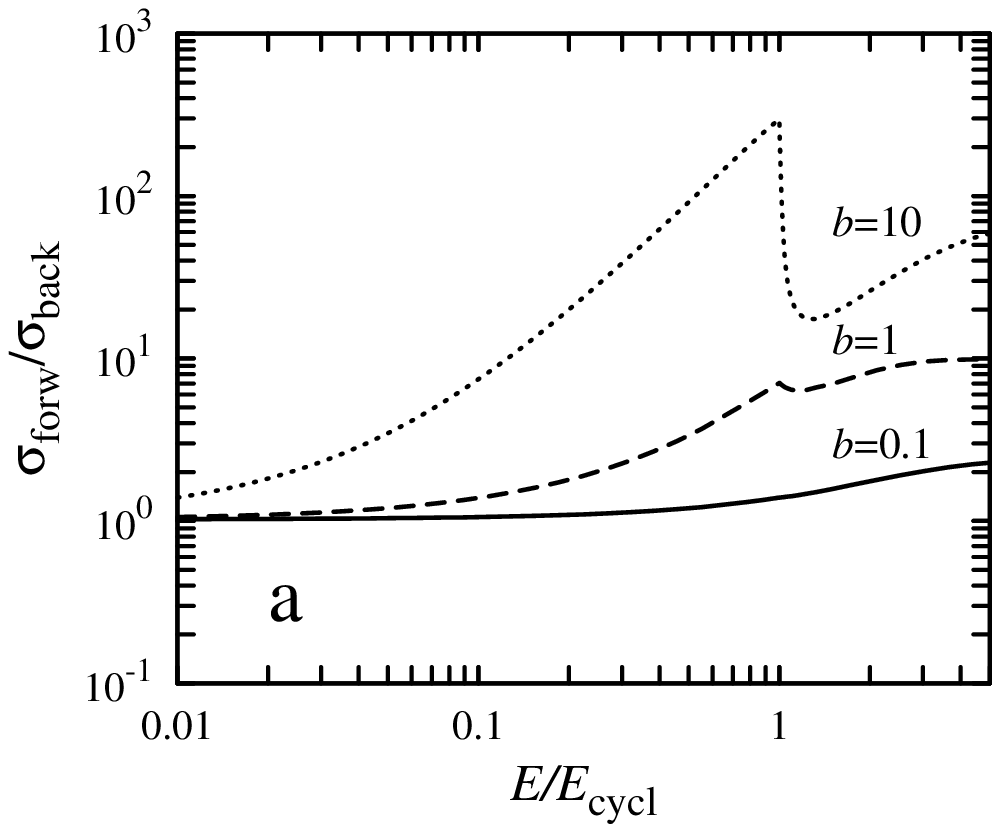}}
\end{minipage}
\hfill
\begin{minipage}{0.49\linewidth}
\center{\includegraphics[angle=0, width=1.\linewidth]{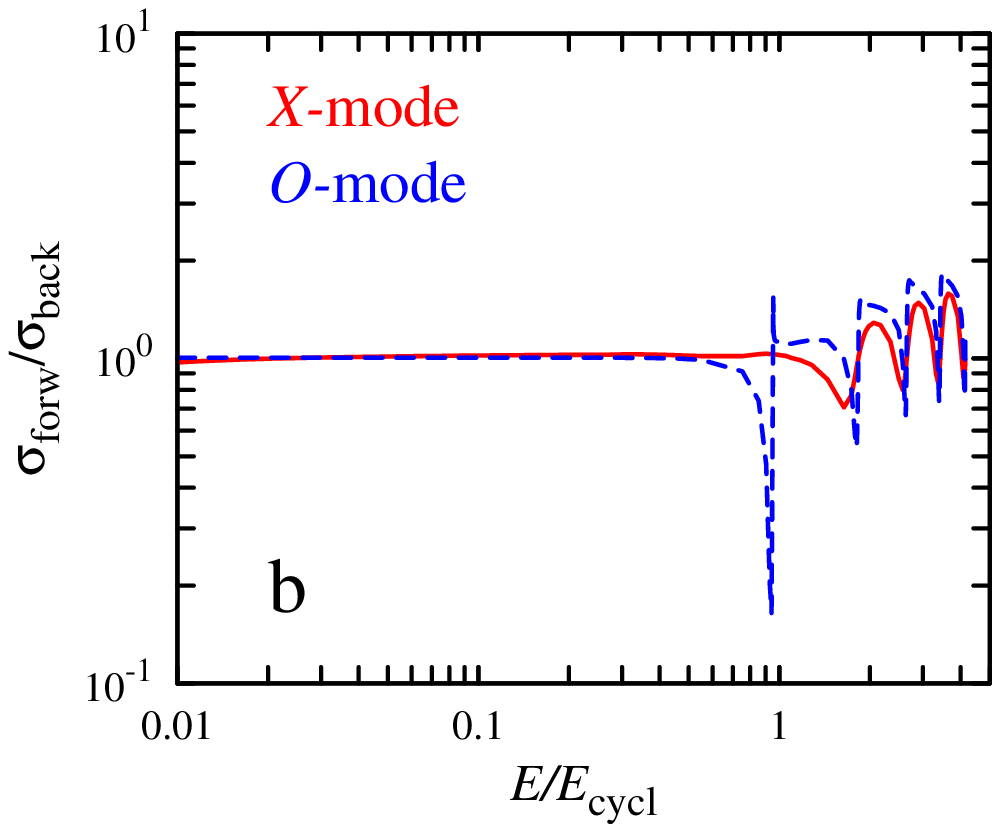}}
\end{minipage}
\caption{The ratio of forward and backward parts of the total scattering cross sections as a function of photon energy (a) for the case of photon initially propagating along ($\theta_i=0$) the $B$-field of different strengths and (b) across the field ($\theta_i=\pi/2$) of given strength $b=0.1$ ($B\simeq 4.4\times 10^{12}$G) for both photon polarizations. Sharp features appear near the resonant energies, where the electrons are able to change their Landau level. The ratio depends strongly on the initial photon momentum direction. The behaviour near the resonant energies also depends on the photon polarization.}
\label{pic:csAssym}
\end{figure}

The photon redistribution over the energies and momentum directions, which is given by differential cross section and redistribution 
function, is not trivial in general case and has to be studied carefully in each particular situation. Additional properties are caused by electron transitions between various Landau levels in a scattering event. Photon redistribution over the directions depends on the initial photon momentum direction, which is a special feature of scattering in the external field, and on the photon energy, which is typical even for the non-magnetic scattering \citep{KN1929}: the scattering indicatrix becomes more elongated in the direction of initial photon momentum as the photon 
energy increases. The scattering in the external magnetic field keeps this regularity but the scattering near the resonant energies
adds additional features (Fig. \ref{pic:csAssym}) corresponding to electron transition between Landau levels: as soon as the photon energy reaches the resonant value, the ratio of forward to backward scattering cross section decrease steeply. It is potentially important for calculation of the radiation pressure resulting from a resonant Compton scattering and particularly important for constructing detailed theory of formation of a beam pattern in X-ray pulsars near the cyclotron energy.

\section{Summary}

Compton scattering of polarized radiation in strong magnetic field is considered. 
A general recipe for calculation of scattering cross section (both differential and total) and $S$-matrix elements based on second order QED perturbation theory is given as well as a recipe for calculation of photon redistribution function over photon energy, momentum and polarization. The presented scheme is adapted both for the scattering by electron with fixed momentum and for the scattering by ensemble of electrons with a given distribution over momentum. A number of calculations in our scheme were simplified analytically. As a result the discussed recipe is sufficiently easy-to-use. Because in our derivation we assume $k=|\vk|c$, the obtained scheme is valid up to magnetic fields of a few hundreds of the Schwinger critical value ($\sim 10^{16}$G), which covers the observed range of neutron stars magnetic field strengths including the extremely high field of magnetars. The scheme is also valid for relatively low magnetic field strength - $10^{6}-10^{9}\,{\rm G}$ - which are typical for white dwarfs, but corresponding calculations with our scheme demand large number of Landau levels which have to be taken into account. The scheme can be used in modelling the atmospheres of neutron stars, where the scattering cross section defines the opacity \citep{2014PhyU...57..735P}. The calculations do not assume any principal restrictions of electron momentum. It gives us a possibility to analyse directly the scattering by moving plasma, which is important for conceptions of X-ray pulsars and accreting neutron stars in general \citep{2007ApJ...654..435B,Poutanen2013,Serber2000}, where Compton scattering governs plasma dynamics in the acceretion channel near the stellar surface \citep{1975A&A....42..311B} and interaction between the radiation and matter in the accretion column for the case of bright X-ray pulsars \citep{Musht2014,1976MNRAS.175..395B}. Our scheme does not contain serious restrictions on the photon energy. The correct Landau level width based on the Sokolov \& Ternov electron wave functions \citep{ST1968,ST1986} is taken into account in general case of Compton scattering for the first time, which generalizes calculations performed earlier by Gonthier et al. \citep{Gont2014}, which were valid for the particular case of initial photon propagating along the magnetic field and ground-to-ground state transition of the electron. The exact spin dependent width of  the levels affect much the resonant scattering cross section of polarized radiation \citep{Gont2014}. Therefore, it has to be taken into account in models describing formation of the cyclotron features in spectra of neutron stars \citep{2008ApJ...672.1127N,2011A&A...532A..76F}.

We have discussed separately the elements of the scattering matrix, which are important for solution of exact relativistic kinetic equation for Compton scattering in a strong $B$-field obtained in our resent work \citep{Mu2012}. It was shown that the $S$-matrix element are real numbers in case when they describe scattering with polarisation changes only.

Potentially important astrophysical results arise from the behaviour of resonant scattering. The resonance position depends on the direction. The stronger the $B$-field, the stronger the dependence (see Fig. \ref{pic:reson} \textit{left}). The position of the fundamental varies by $\approx20\,\%$ for $B\sim10^{13}\,{\rm G}$ and even more for higher field strength. It can be used in diagnostics of X-ray pulsars since this effect would partly define the changes of the cyclotron absorption line position during the pulse period \citep{2015MNRAS.448.2175L}. The ratio of the energies of first and second resonances $k_{\rm res}^{(2)}/k_{\rm res}^{(1)}$ is also depend on the direction in a strong magnetic field (see Fig. \ref{pic:reson} \textit{right}), and it can cause the change of the ratio of cyclotron line energies during pulse period \citep{2015MNRAS.448.2175L} and nonequidistance of the cyclotron line harmonics, which was observed in spectra of X-ray pulsars \citep{TsL2006}. The effect is also causes the variations of scattering cross section with the angle even for the case of $X$-mode photons (see Fig. \ref{pic:scSigAng}). It is particularly important for radiation transfer and radiation pressure calculation in case of high $B$-field, since the opacity would strongly depend on directions. It was pointed that the photon redistribution over directions changes as soon as the initial photon energy crosses the resonant value (see Fig. \ref{pic:csAssym}). It is potentially important for the formation of a beam pattern of X-ray pulsars near the cyclotron line. 

The presented scheme of calculation provides a ground for investigation of radiation transfer in strongly magnetised plasma. It can be readily applied to astrophysical problems, principally for the models of spectrum formation in strongly magnetized neutron stars, calculation of radiation pressure in strong $B$-field and modelling of X-ray pulsar beam pattern all over the spectrum. In this way, the presented scheme is extremely relevant to further investigation of strongly magnetized neutron stars.

\acknowledgments

This study was supported by Magnus Ehrnrooth foundation grants (A.M.), the Saint-Peterburg State University grants 6.0.22.2010, 6.38.669.2013, 6.38.18.2014 (D.N.), the Academy of Finland grant 268740 (J.P.). We are grateful to Dmitry Yakovlev, Valery Suleimanov, Dmitry Rumyantsev and Sergey Tsygankov for a number of useful comments.

\bibliography{allbib}

\appendix

\section{Longitudinal transformation of electron and photon momenta}

We are focusing here on the longitudinal Lorentz transformation of the timelike component of the 4-momentum $p_0$, i.e. particle's energy, and the $z$-component of the momentum $p_z$, which corresponds to the particle momentum along the magnetic field. The general form of the longitudinal transformation is
\be \label{eq:Lorenz}
p_0'=(p_0-\beta\,p_z)/\sqrt {1-\beta^2},\quad p'_z=(p_z-\beta\,p_0)/\sqrt {1-\beta^2},
\ee
where $\beta$ is the velocity between the reference frames along the magnetic field in units of speed of light. The transformation (\ref{eq:Lorenz}) can be rewritten in another form using parameter  $\chi$, which satisfies the relation $\beta=\tanh\chi$. Then 
\be \label{eq:Lorenchi}
p_0'=p_0\cosh\chi-p_z\sinh\chi,\quad p_z'=p_z\cosh\chi-p_0\sinh\chi.
\ee
Thus the photon energy and the longitudinal momentum are transformed as follows:
\be \label{eq:kkcoschsh}
k'=k(\cosh\chi-\sinh\chi\cos\theta),\quad k'\cos\theta'=k(\cos\theta\cosh\chi-\sinh\chi).
\ee
The transformation of the angle between the $B$-field direction and the photon momentum is given by the relations
\be \label{eq:cossinchsh}
\cos\theta'=\frac {\cos\theta\cosh\chi-\sinh\chi}{\cosh\chi-\sinh\chi\cos\theta},\,\,
\sin\theta'=\frac {\sin\theta}{\cosh\chi-\sinh\chi\cos\theta}.
\ee
The electron energy $E_n$ and momentum along the magnetic field $p_z$ are transformed according to (\ref{eq:Lorenz}):
\be \label{eq:RZchsh}
E'_n=E_n\cosh\chi-p_z\sinh\chi,\quad p_z'=p_z\cosh\chi-E_n\sinh\chi.
\ee
Using relation $\d E_n=p_z\d p_z/E_n$ we get:
\be \label{eq:dZstdZ}
\d p_z'=\d p_z\cosh\chi-\d E_n\sinh\chi=\left(\cosh\chi-\frac {p_z}{E_n}\sinh\chi\right)\d p_z=
\frac {\d p_z}{E_n}(E_n\cosh\chi-p_z\sinh\chi)=\frac {E_n'}{E_n}\d p_z.
\ee
Therefore the ratio ${\d p_z}/{E_n}$ is conserved under longitudinal Lorentz transformation.

If photon energy $k$ and momentum along the field $k_z$ are given at the laboratory reference frame, where the electron momentum along the field is $p_z$, then the photon energy $k'$ and momentum $k'_z$ in the electron reference frame (where $p_z=0$) are
\be \label{rq:KRZ}
k'=\frac {k}{\sqrt{E_n^2-p_z^2}}(E_n-p_z\cos\theta),\quad k_z'=\frac {k}{\sqrt{E_n^2-p_z^2}} (E_n\cos\theta-p_z).
\ee
The angle $\theta'$ between the photon momentum and magnetic field direction satisfies following relations:
\be \label{eq:aberrat}
\cos\theta'=\frac {E_n\cos\theta-p_z}{E_n-p_z\cos\theta},\quad
\sin\theta'=\frac {\sqrt{E_n^2-p_z^2}\sin\theta}{E_n-p_z\cos\theta}.
\ee

\section{Landau level natural width}
\label{sec:LandauLevelWidth}

Landau level natural width for the particular case of $p_{z,i}=0$ is defined as a sum of the partial widths:
\be
\Gamma_n^\pm=\sum\limits_{n'<n}\Gamma_{nn'}^\pm.
\ee
The general expression for the partial width was obtained by Herold et al. \citep{Her1982} for transition between arbitrary Landau levels but for zero initial electron momentum $p_{z,i}=0$:
\beq
& \strut\disp
\Gamma_{nn'}^\pm=\frac{r_{\rm e}}{2}\int\limits_{0}^{\pi/2}\d\theta\frac{k\sin\theta}{E_n\sqrt{E^2_{n}-2(n-n')b\sin^2\theta}}
& \nonumber \\
& \strut\disp 
\times \left\{\left[(E_n\mp 1)(E_n\pm 1-k)I^2_{n,n'}(u)+(E_n\pm 1)(E_n\mp 1-k)I^2_{n-1,n'-1}(u)\right]\sin^2\theta\right.
& \nonumber \\
& \strut\disp 
+\left[(E_n\pm 1)(E_n\mp 1-k)I^2_{n-1,n'}(u)+(E_n\mp 1)(E_n\pm 1-k)I^2_{n,n'-1}(u)\right](1+\cos^2\theta)
& \nonumber \\
& \strut\disp 
+2k\sqrt{2nb}\left[I^2_{n,n'}(u)I^2_{n-1,n'}(u)+I^2_{n,n'-1}(u)I^2_{n-1,n'-1}(u)\right]\sin\theta\cos^2\theta
& \nonumber \\
& \strut\disp 
+\left. 4b\sqrt{nn'}\left[I^2_{n-1,n'}(u)I^2_{n,n'-1}(u)+I^2_{n,n'}(u)I^2_{n-1,n'-1}(u)\right]\sin^2\theta\right\},
&
\eeq
where $E_n=\sqrt{1+2nb}$ is the electron energy, $k=\left[E_n-(E^2_n-2(n-n')b\sin^2\theta)^{1/2}\right]/\sin^2\theta$ is the energy of a photon emitted at the angle $\theta$ due to the electron transition $n\longrightarrow n'$,
\be
I_{n,n'}(u)=(-1)^n(n!n'!)^{-1/2}\exp[u/2]u^{(n-n')/2}\frac{\partial^n}{\partial u^n}\left(u^{n'}\exp[-u]\right)
\ee
and $u=(k^2\sin^2\theta)/2b$ \citep{1977FCPh....2..203C}. The functions $I_{n,n'}(u)$ can be constructed using the associated Laguerre polynomials $L_n^\alpha(x)$:
\be
I_{n,n'}(u)=(-1)^{n}(n!/n'!)^{1/2}\exp[-u/2]u^{(n'-n)/2}L_{n}^{n'-n}(u),\quad 
L_n^\alpha(x)\equiv \frac{e^x x^{-\alpha}}{n!}\frac{\d^n}{\d x^n}\left[e^{-x}x^{n+\alpha}\right].
\ee

\begin{figure}
\begin{minipage}{0.49\linewidth}
\center{\includegraphics[angle=0, width=1.\linewidth]{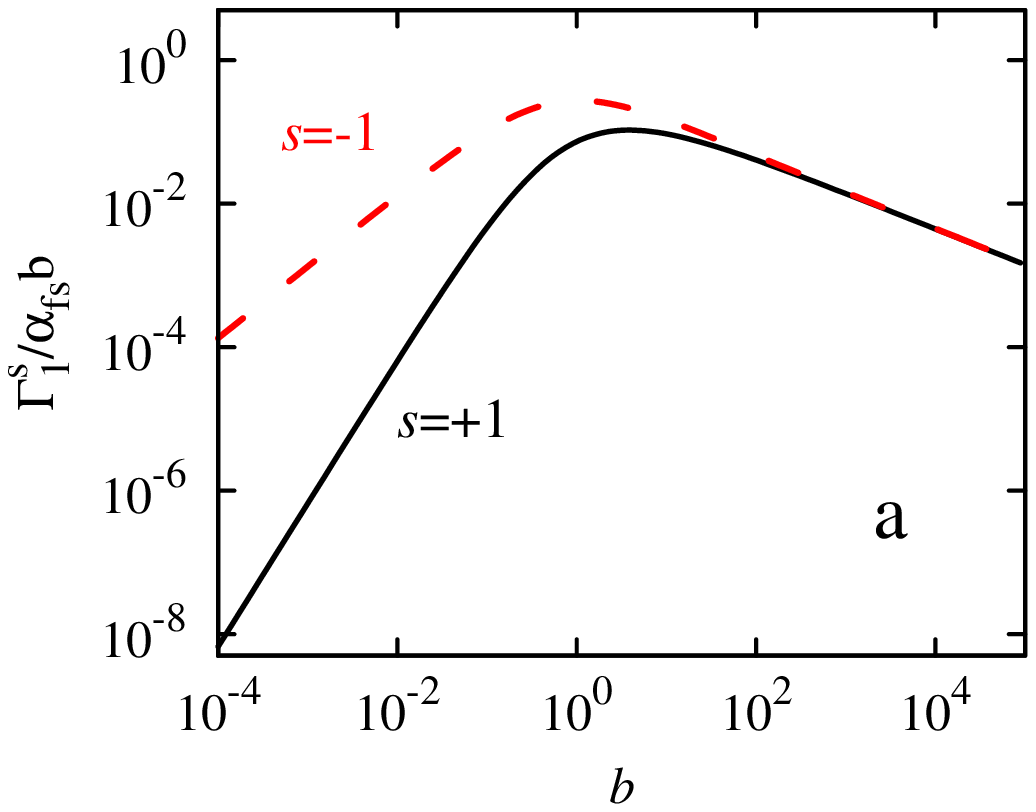}}
\end{minipage}
\hfill
\begin{minipage}{0.49\linewidth}
\center{\includegraphics[angle=0, width=1.\linewidth]{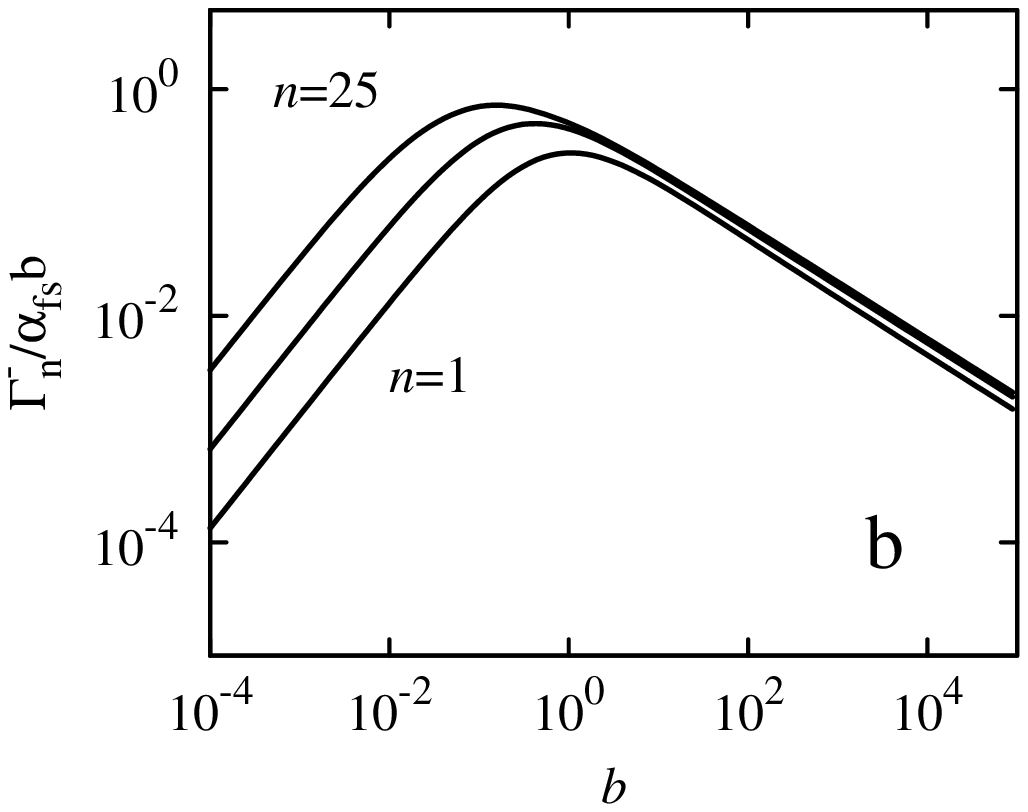}}
\end{minipage}
\caption{Spin dependent Landau level width in the electron rest frame (in units of $\alpha_{\rm fs}b$) as a function of dimensionless magnetic field strength $b=B/B_{\rm cr}$ is given for the first Landau level (a) and for the 1st, 5th and 25th Landau levels of spin state $s=-1$ (b).}
\label{pic:LLw}
\end{figure}

Compact approximate expressions for $\Gamma^\pm_{n}$ and $\Gamma^{\pm}_{nn'}$ for the particular cases of $nb\ll 1$ (non-relativistic limit), $b^{-1}\ll n\ll b^{-3}$ (ultrarelativistic quasi-classical limit)  and $n\gg b^{-3}$ (ultrarelativistic quantum limit) were provided by Pavlov et al. \citep{Pavlov1991}. The Landau level widths for the case of nonzero momentum of the electron along the field $p_{z,i}\neq 0$ can be obtained from those expressions for $p_{z,i}=0$ by Lorentz transformation \citep{Her1982}: 
$\Gamma_n^\pm(p_{z})=\Gamma_n^\pm \sqrt{1+2bn}/E_{n}(p_{z})$.

\begin{figure}
\begin{minipage}{0.49\linewidth}
\center{\includegraphics[angle=0, width=1.\linewidth]{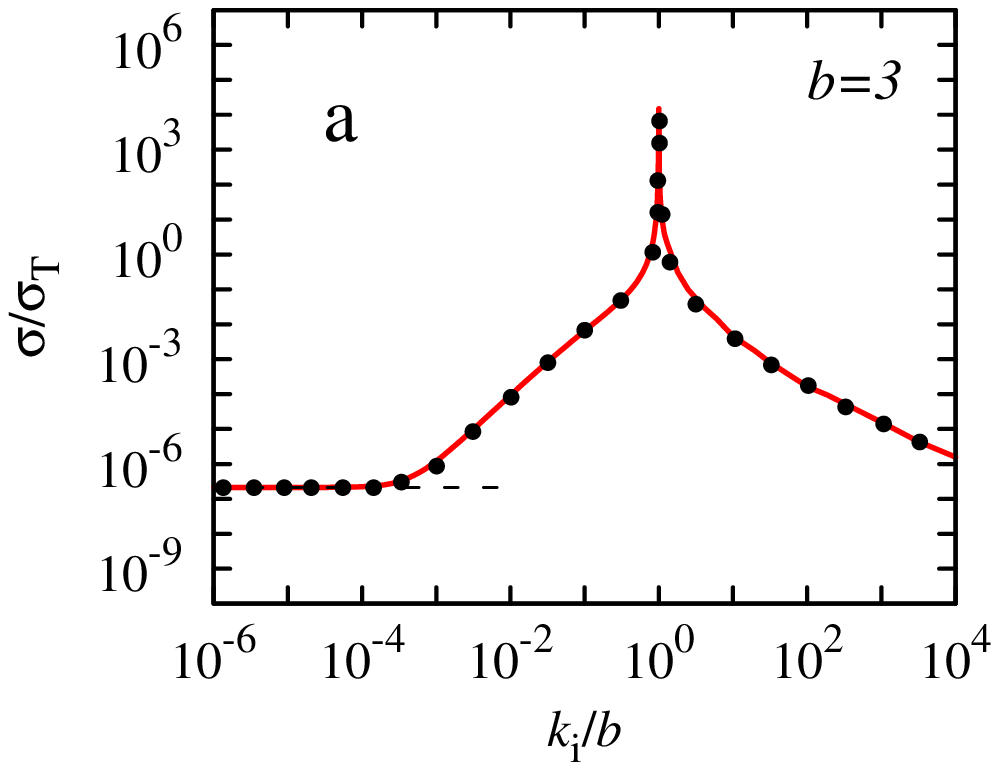}}
\end{minipage}
\hfill
\begin{minipage}{0.49\linewidth}
\center{\includegraphics[angle=0, width=1.\linewidth]{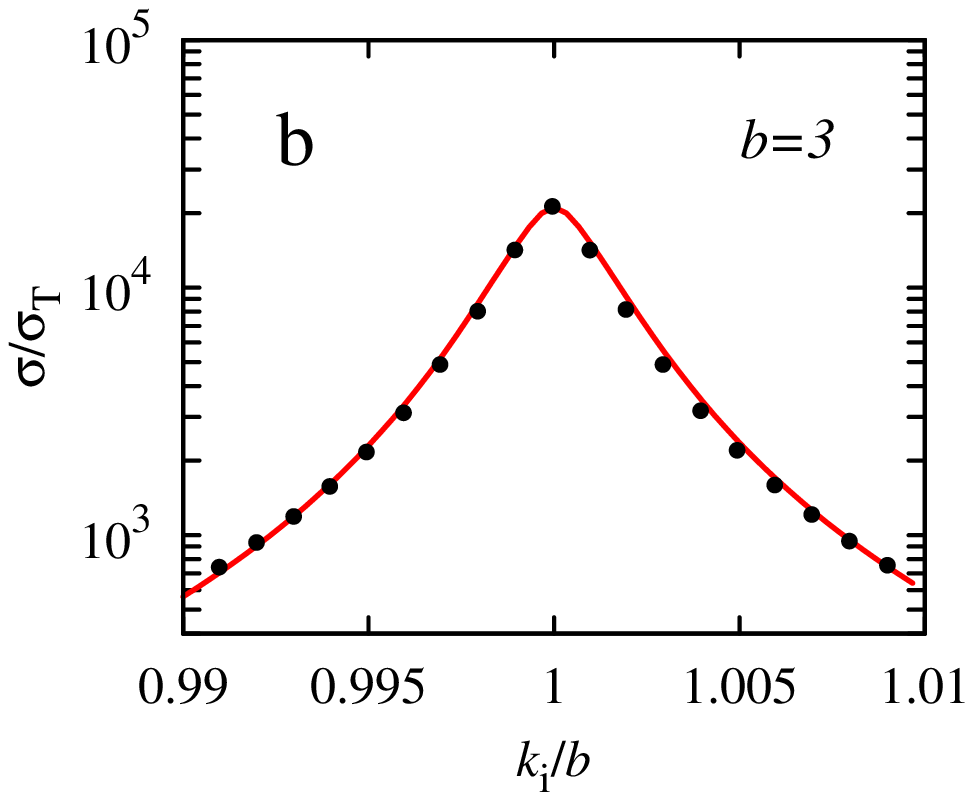}}
\end{minipage}
\caption{Compton scattering cross section calculated using spin-dependent Landau level width for the particular case of ground-to-ground state transition is given by the red solid lines in a wide photon energy range (a) and for photon of energy close to the fundamental (b). The magnetic field strength $b=B/B_{\rm cr}=3$ and initial angle between the field direction and photon momentum $\theta_i=0$. Black dots represent calculations performed by Gonthier et al. \citep{Gont2014} for the same conditions. The Landau level width affects strongly the cross section at low energies, where the level width becomes comparable to the photon energy and the cross section saturates at a small constant value (\ref{eq:CSappLLw}) given by the black dashed line (a). The level width also affects strongly the cross section at the resonance energies (b).}
\label{pic:CSwLLw}
\end{figure}

\begin{figure}
\begin{minipage}{0.55\linewidth}
\center{\includegraphics[angle=0, width=1.\linewidth]{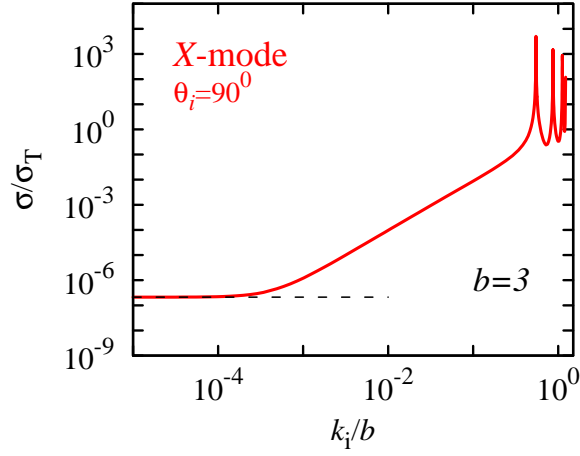}}
\end{minipage}
\caption{Scattering cross section for the $X$-mode photons of initial angle $\theta_i=90^{\circ}$ in a wide range of initial photon energies. The magnetic field strength $b=B/B_{\rm cr}=3$. At low energies the cross section saturates at a small constant value (\ref{eq:CSappLLw}) given by the black dashed line similarly to the case of photons propagating along the $B$-field (see Fig.\,\ref{pic:CSwLLw}a).}
\label{pic:CSwLLw2}
\end{figure}
	
Cyclotron decay rates for transition to the ground state and arbitrary initial electron momentum $p_{z,i}$ were obtained by Latal \citep{1986ApJ...309..372L}. The simplified expressions were introduces by Baring et al. \citep{2005ApJ...630..430B}. Although the resonance line widths involve infinite sums over Landau levels, in the case of fundamental resonance the sum is dominated by the $n=1$ state. The width of this state is equal to the $n\longrightarrow 0$ cyclotron decay rate. As a result, the fundamental line width can be well approximated by the particular cyclotron rate obtained by Latal \citep{1986ApJ...309..372L,2005ApJ...630..430B}. For the case of $b\gg 1$ cyclotron transitions to the ground Landau level dominate \citep{1987ApJ...319..939H} and the cyclotron decay rate for $n\longrightarrow 0$ transitions approximate well the widths of excited states.

Landau level natural width becomes crucially important at resonant photon energies (see Section \ref{sec:ResonReg}) and at energies well below the cyclotron energy, when the initial photon energy becomes comparable to the Landau level width \citep{Gont2014}. If $k_i\ll \Gamma$ the cross section for the photons propagating along the magnetic field saturates at a small value 
\be\label{eq:CSappLLw}
\sigma\approx \sigma_{\rm T}\Gamma^2 b^{-2}(1+2b)^{-1}.
\ee 
The same happens with photons of $X$-mode propagating in any direction (see Fig.\,\ref{pic:CSwLLw2}).

\section{Set of used matrices and useful relations}
\label{sec:DirMatr}
In this section we present the matrices which we use in our calculations. In general we are following the standard designations 
\citep{BSh1959,BLP1971,Peskin1995}. 


We use a set of three $2\times 2$ Pauli matrices, $\sigma_1,\,\sigma_2,$ and $\sigma_3$, which are Hermitian and unitary, in their standard designation \citep{Peskin1995}. $I$ is a unity $2\times 2$ matrix. We also use the following combinations of Pauli matrices:
$
\sigma^{\pm}=(I\pm\sigma_3)/2,\,\,\,
\sigma_{\pm}=(\sigma_1\pm i\,\sigma_2)/2.
$

The gamma (Dirac) matrices which compose the 4-dimensional vector $\ugam=\{\gamma^0,\gamma^1,\gamma^2,\gamma^3\}$
could be expressed via the $2\times 2$ Pauli matrices:
\be \label{eq:gams}
\gamma^0=\gamma_0=\left(\begin {array}{cc} I & 0 \\ 0 & -I 
\end {array} \right),\,\,\,
\gamma^{i}=-\gamma_{i}=\left(\begin {array}{cc} 0 & \sigma_{i} \\ -\sigma_{i} & 0
\end {array} \right).
\ee
We also introduce matrices $\uD=\{D^0,\textbf{D}\}$, where:
\be \label{eq:Ds}
D^0=\left(\begin {array}{cc} 0 & -I \\ I & 0 \end {array}\right),\,\,\,
D^{i}=\left(\begin {array}{cc} -\sigma_{i} & 0 \\ 0 & \sigma_{i}
\end {array}\right),
\ee
and 3-dimensional vectors of matrices $\valp$ and $\vSig$:
\be \label{eq:valpSig}
\alpha_{i}=\left(\begin {array}{cc} 0 & \sigma_{i} \\ \sigma_{i} & 0
\end {array}\right),\,\,\,
\Sigma_{i}=\left(\begin {array}{cc} \sigma_{i} & 0 \\ 0 & \sigma_{i}
\end {array}\right).
\ee
Let us designate the unity matrix $4\times4$ with 1, and the product of four matrices with $\gamma_5$: 
\be \label{eq:gam5}
\gamma_5=-\gamma^5=i\,\gamma^0\,\gamma^1\,\gamma^2\,\gamma^3=
\left(\begin {array}{cc} 0 & I \\ I & 0 \end {array}\right).
\ee

We also use the following linear combination of the matrices:
\be \label{eq:Siglin}
\Sigma^{\pm}=(I\pm\Sigma_3)/2,\,\,
\Sigma_{\pm}=(\Sigma_1\pm i\,\Sigma_2)/2,\,\,
\alpha^{\pm}=(\gamma_5\pm\alpha_3)/2,\,\,
\alpha_{\pm}=(\alpha_1\pm i\,\alpha_2)/2,
\ee
\be \label{eq:Dlin}
D^{\pm}=(\gamma^0\mp D^3)/2,\,
D_{\pm}=-(D^1\pm i\,D^2)/2,\,
\gamma^{\pm}=(-D^0\pm\gamma^3)/2,\,
\gamma_{\pm}=(\gamma^1\pm i\,\gamma^2)/2.
\ee
These matrices compose the set of 16 linearly independent $4\times4$ matrices. 
They could be expressed via $2\times 2$ matrices in the following way:
\beq \label{eq:lincomb1}
& \Sigma^{\pm}=\left(\begin {array}{cc} \sigma^{\pm} & 0 \\ 0 & \sigma^{\pm} 
\end {array} \right),\,\,\,
\Sigma_{\pm}=\left(\begin {array}{cc} \sigma_{\pm} & 0 \\ 0 & \sigma_{\pm} 
\end {array} \right),\,\,\,
\alpha^{\pm}=\left(\begin {array}{cc}  0 & \sigma^{\pm} \\ \sigma^{\pm} & 0
\end {array} \right),\,\,\,
\alpha_{\pm}=\left(\begin {array}{cc}  0 & \sigma_{\pm} \\ \sigma_{\pm} & 0
\end {array} \right), & \\ \label{eq:lincomb2}
& D^{\pm}=\left(\begin {array}{cc} \sigma^{\pm} & 0 \\ 0 & -\sigma^{\pm} 
\end {array} \right),\,\,\,
D_{\pm}=\left(\begin {array}{cc} \sigma_{\pm} & 0 \\ 0 & -\sigma_{\pm} 
\end {array} \right),\,\,\,
\gamma^{\pm}=\left(\begin {array}{cc}  0 & \sigma^{\pm} \\ -\sigma^{\pm} & 0
\end {array} \right),\,\,\,
\gamma_{\pm}=\left(\begin {array}{cc}  0 & \sigma_{\pm} \\ -\sigma_{\pm} & 0
\end {array} \right). &
\eeq

The Dirac matrices are determined by relations of anticommutativity. For the 4-vectors of matrices they are
\be \label{eq:rel4}
\gamma^\mu\gamma^\nu+\gamma^\nu\gamma^\mu=2g^{\mu\nu},\quad
\gamma_5\ugam+\ugam\gamma_5=0,\quad
D^\mu D^\nu+D^\nu D^\mu=-2g_{\mu\nu},\quad \gamma_5\uD+\uD\gamma_5=0,
\ee
and for the 3-vectors of matrices the relations are
\be \label{eq:rel3}
\alpha_k\alpha_j+\alpha_j\alpha_k=2\delta_{kj},\quad
\Sigma_k\Sigma_j+\Sigma_j\Sigma_k=2\delta_{kj},\quad
D^k\Sigma_j+\Sigma_jD^k=-2\gamma^0\delta_{kj}.
\ee
Useful commutative relations are:
\be \label{eq:commuts}
D^\mu\gamma^\nu-\gamma^\nu D^\mu=2\gamma_5g^{\mu\nu},\quad
\gamma^k\alpha_j-\alpha_j\gamma^k=2\gamma^0\delta_{kj}.
\ee
The useful dot products of the 4-vectors ($\unl{a}\,\unl{b}\equiv a^0b^0-\sum_{i=1}^3 a^i b^i$) of matrices are
\be \label{eq:ugamD}
\ugam\,\ugam=4,\quad 
\uD\,\uD=-4,\quad
\ugam\uD=-\uD\ugam=-4\gamma_5
\ee
and for the 3-vectors ($\textbf{ab}\equiv\sum_{i=1}^3 a^i b^i$) of matrices are
\beq \label{eq:vgamD}
& \vgam\vgam=-3,\quad \vgam\valp=-\valp\vgam=3\gamma^0,\quad \vgam\vSig=
\vSig\vgam=-3D^0,\quad \vgam\textbf{D}=-\textbf{D}\vgam=3\gamma_5, & \\
& \valp\valp=3,\quad \valp\vSig=\vSig\valp=3\gamma^0,\quad 
\valp\textbf{D}=-\textbf{D}\valp=-3D^0, & \\ 
& \vSig\vSig=3,\quad \textbf{D}\textbf{D}=3,\quad \vSig\textbf{D}=\textbf{D}\vSig=3\gamma^0. &
\eeq

\section{Electron in the external magnetic field}
\label{sec:ElectronInMF}

In this section we discuss the description of an electron in the external magnetic field which we use
in this paper. The different ways of electron description in such case are also discussed in literature
\citep{ST1986,Her1982,M1992}.

\subsection{Dirac equation}

The electron is described by the Dirac equation, which has to be written for the case of external magnetic field. 
Let us choose 4-vector of potential in the Landau gauge: 
$\uA_\e=\{0,\vA_\e\}$, where $\vA_\e=B_\e\,(0,x,0)$. Then the required solutions $\Psi$ satisfy the
equation:
\be \label{eq:eqDPsi}
\left(\hat {p}+\frac {e}{c}\hat {A}_\e-m\,c\right)\,\Psi=0,
\ee
where $\hat {A}_\e=\uA_\e\ugam=-B_\e\,x\gamma^2$, $\gamma^2$ is one of the Pauli matrices (\ref{eq:gams}) and $\hat {p}=\up\,\ugam=i\,\hbar\,\unb\ugam$.

Let us use relativistic quantum system of units and find the solution in the following form:
$\Psi=\exp {[i(-Et+p_{z}z+p_{y}y)]}\,\psi$. It is useful to change the variables: $x=u-p_{y}/b$. Then Dirac equation (\ref{eq:eqDPsi}) takes form
\be \label{eq:eqDpsi}
\left(E\,\gamma^0+i\,\frac {\d}{\d u}\gamma^1-p_{z}\,\gamma^3-
b\,u\gamma^2-1 \right)\,\psi(u)=0.
\ee
If it is multiplied by $i\gamma^1$, then we get the ordinary system of differential equations:
\be \label{eq:eqsnorm}
\left(\frac {\d}{\d u}-i\,E\,\alpha_1-b\,u\,\Sigma_3+p_{z}\,\Sigma_2-
i\,\gamma^1\right)\,\psi(u)=0.
\ee

\subsection{From the system of equations to second order differential equations}

Let designate the components of the vector which we want to find:
$\psi(u)=(\psi_1,\psi_2,\psi_3,\psi_4)^{\rm T}$, and rewrite the ordinary system of differential 
equations (\ref{eq:eqsnorm}) in details:
\be \label{eq:normeqs}
\left\{\begin {array}{lllll} \disp
{\d\psi_1}/{\d u}= b u\psi_1 +p_{z}i\psi_2 +0 +i(E+1)\psi_4 \\
\disp {\d \psi_2}/{\d u}=-p_{z}i\psi_1 -b u\psi_2 +i(E+1)\psi_3 +0 \\
\disp {\d \psi_3}/{\d u}= 0 +i(E-1)\psi_2 +b u\psi_3 +p_{z}i\psi_4 \\
 \disp {\d \psi_4}/{\d u}= i(E-1)\psi_1 +0 -p_{z}i\psi_3 -b u\psi_4
\end {array}\right. .
\ee
Then we can find equations for each function in (\ref{eq:normeqs}):

1) The case of $\psi_1=0$ gives an equation for $\psi_3$. Using the designations: $\zeta\equiv(E+1)\psi_3$ and $a\equiv E^2-1-p_{z}^2$, we get:
$
\psi_1=0,\quad p_{z}\psi_2+(E+1)\psi_4=0,\quad
\psi_2'=-b u\psi_2+i\zeta,\quad \zeta'=ia\psi_2+b u\zeta.
$
Therefore:
\be \label{eq:eqpsi}
\zeta''=(b^2u^2+b-a)\zeta.
\ee

2) The case of $\psi_2=0$ gives a solution for $\psi_4$. Defining $\mu\equiv (E+1)\psi_4$, we get 
$\psi_2=0$, $-p_{z}\psi_1+(E+1)\psi_3=0$, $\psi_1'=b u\psi_1+i\mu$, $\mu'=ia\psi_1-b u\mu$.
Therefore:
\be \label{eq:eqphi}
\mu''=(b^2u^2-b-a)\mu.
\ee

3) The case of $\psi_3=0$ gives us a solution for $\psi_1$. Defining $\eta\equiv(E-1)\psi_1$, we get 
$\psi_3=0$, $(E-1)\psi_2+p_{z}\psi_4=0$, $\psi_4'=-b u\psi_4+i\eta$, $\eta'=ia\psi_4+b u\eta$.
Here we get the same equation as in a first case (\ref{eq:eqpsi}):
$\eta''=(b^2u^2+b-a)\eta.$

4) The case of $\psi_4=0$ gives us a solution for $\psi_2$. Defining $\kappa\equiv(E-1)\psi_2$ and using similar designations as in the second case we get $\psi_4=0$, $(E-1)\psi_1-Z\psi_3=0$, $\psi_3'=b u\psi_3+i\kappa$, $\kappa'=ia\psi_3-b u\kappa$.
And we get the same equation as in a second case: $\kappa''=(b^2u^2-b-a)\kappa$.

Thus the system of equations (\ref{eq:normeqs}) is reduced to the pair of equations of the same form: (\ref{eq:eqpsi}) and (\ref{eq:eqphi}).
Both of them can be transformed to the equation of quantum harmonic oscillator. Its solutions are well known and enumerated with integer numbers $n\geq 0$:
\be \label{eq:eqosc}
\frac {1}{2}\left[-\DR {^2\phi_n(\xi)}{\xi^2}+\xi^2\phi_n(\xi)\right]=
\left(n+\frac {1}{2}\right)\phi_n(\xi).
\ee
The eigen functions could be written via the Hermite polynomials: $\phi_n(\xi)=\pi^{-1/4}(2^nn!)^{-1/2}e^{-\xi^2/2}H_n(\xi)$. Thus, we find that the motion of electrons is quantized and they occupy Landau levels.

The eigen functions form orthonormalized series. The expressions for the derivative take the form:
$$
\phi_n'(\xi)=\sqrt {2n}\phi_{n-1}(\xi)-\xi\phi_n(\xi),\,\,
\phi_{n-1}'(\xi)=-\sqrt {2n}\phi_n(\xi)+\xi\phi_{n-1}(\xi).
$$

Our solutions will be expressed through the functions $\chi_n(u)$, which are defined by harmonic oscillator eigen functions $\phi_n(\xi)$
and comply with the relations:
\beq \label{eq:chin}
& \strut\disp \chi_n(u)=b^{1/4}i^n\phi_n(b^{1/2}u),\,\,
\frac {1}{2}\left[-\frac {1}{b}\frac {\d^2\chi_n(\xi)}{\d\xi^2}+
b\xi^2\chi_n(\xi)\right]=
\left(n+\frac {1}{2}\right)\chi_n(\xi), & \\ \label{eq:chinn1}
& \strut\disp \chi_n'(u)=ib_n\chi_{n-1}(u)-b u\chi_n(u),\,\,
\chi_{n-1}'(u)=ib_n\chi_n(u)+b u\chi_{n-1}(u),\,\,b_n=\sqrt {2b n}. &
\eeq
Functions $\chi_n(u)$ are normalized:
$\int_{-\infty}^\infty\chi^*_n(\xi)\chi_{n'}(\xi)\d\xi=\delta_{nn'}$.

\subsection{Solution of the system of equations}

The solutions of the second order equations (\ref{eq:eqpsi},\ref{eq:eqphi}) give us a solution of
the system of the equations (\ref{eq:normeqs}). Let us enumerate the solutions with the upper index $(l)$
and gather them into the matrix $\left(\psi_j^{(l)}(u)\right)$:
\be \label{eq:soleqs}
\psi=\left(\psi_j^{(l)}(u)\right)=\left(\begin {array}{cccc}
(E+1)\chi_{n-1} & 0 & p_{z}\chi_{n-1} & b_n\chi_{n-1} \\
0 & (E+1)\chi_n & b_n\chi_n & -p_{z}\chi_n \\
p_{z}\chi_{n-1} & b_n\chi_{n-1} & (E-1)\chi_{n-1} & 0 \\
b_n\chi_n & -p_{z}\chi_n & 0 & (E-1)\chi_n
\end {array}\right).
\ee

However, these solutions are linearly dependent:
$(E-1)\psi^{(1)}=p_{z}\psi^{(3)}+b_n\psi^{(4)}$, 
$(E-1)\psi^{(2)}=b_n\psi^{(3)}-p_{z}\psi^{(4)}$.
In order to get four independent solutions one have to use the ones with the negative energy 
$E=\pm E_n$, $E_n=\sqrt{1+b_n^2+p_{z}^2}$, which correspond to the positrons.
Let us write down the solutions. Two of them correspond to the electrons and have the form:
\be \label{eq:solel}
\Psi_{nj}^{+}(x,y,z,p_{y},p_{z})=\left(\frac {E_n+1}{2E_n}\right)^{1/2}v_{nj}^{+}(p_{z},u)
e^{-i(E_nt-p_{y}y-p_{z}z)},\,\,j=1,2,
\ee
where
$$v_{n1}^{+}(p_{z},u)=\left(\begin {array}{c}
\chi_{n-1}(u) \\ 0 \\ p_{z}\chi_{n-1}(u)/(E_n+1) \\ b_n\chi_n(u)/(E_n+1)
\end {array}\right),\qquad
v_{n2}^{+}(p_{z},u)=\left(\begin {array}{c}
0 \\ \chi_n(u) \\ b_n\chi_{n-1}(u)/(E_n+1) \\ -p_{z}\chi_n(u)/(E_n+1)
\end {array}\right).$$
And two of them correspond to the positron states:
\be \label{eq:solpos}
\Psi_{nj}^{-}(x,y,z,p_{y},p_{z})=\left(\frac {E_n+1}{2E_n}\right)^{1/2}v_{nj}^{-}(p_{z},u)
e^{i(E_nt+p_{y}y+p_{z}z)},\,\,j=1,2,
\ee
where
$$v_{n1}^{-}(p_z,u)=\left(\begin {array}{c}
-p_{z}\chi_{n-1}(u)/(E_n+1) \\ -b_n\chi_n(u)/(E_n+1) \\ \chi_{n-1}(u) \\ 0 
\end {array}\right),\qquad
v_{n2}^{-}(p_{z},u)=\left(\begin {array}{c}
-b_n\chi_{n-1}(u)/(E_n+1) \\ p_{z}\chi_n(u)/(E_n+1) \\ 0 \\ \chi_n(u)
\end {array}\right)$$
and $u=x+p_{y}/b$.

The wave functions could be also presented in the following form:
\be \label{eq:Psipos}
\Psi_{nj}^{-}(x,y,z,-p_{y},-p_{z})=\left(\frac{E_n+1}{2E_n}\right)^{1/2}
v_{ns}^{-}(-p_{z},u)e^{i(E_nt-p_{y}y-p_{z}z)},\,\,s=1,2,
\ee
where $u=x-p_{y}/b$.
In case of $n=0$ two solutions vanish: $v_{01}^{+}(p_{z},u)=v_{01}^{-}(p_{z},u)=0$.

\subsection{The solutions for definite helicity}

Let as find now the solutions in a form when they are eigenvectors of the helicity operators
$\tilde{S}$ and $\mu$ (non self-conjugated and self-conjugated correspondingly) \citep{BSh1959}. 
They would be the linear combinations of the solutions with indexes $s=1,2$.
The helicity operator $\tilde{S}$ acts on the 4-vectors only and it does not act on the functions $\chi$.
Therefore these functions are multiplier factors in front of the eigenvectors of the operator $\tilde{S}$: 
\be\label{eq:Ui}
U_1(p_z)=\left(\begin {array}{c}
E_n+s_n \\ 0 \\ p_z \\ 0 
\end {array}\right),\quad
U_2(p_z)=\left(\begin {array}{c}
0 \\ E_n+s_n \\ 0 \\ -p_z 
\end {array}\right),\quad
U_3(p_z)=\left(\begin {array}{c}
p_z \\ 0 \\ E_n+s_n \\ 0 
\end {array}\right),\quad
U_4(p_z)=\left(\begin {array}{c}
0 \\ -p_z \\ 0 \\ E_n+s_n 
\end {array}\right),
\ee
where $E_n=\sqrt{1+2nb+p_z^2}$ is the particle energy and $s_n=\sqrt{1+2bn}$. Thus it is necessary to consider four linear combinations.

1) For the electron with the helicity $+1$ the following relation could be written down: 
$$
C_1v^{+}_{n1}(p_{z},u)+C_2v^{+}_{n2}(p_{z},u)=v_{n+}^{+}(p_{z},u)=
\alpha_1U_1\chi_{n-1}+\alpha_2U_4\chi_n.
$$
And therefore one finds out the relations for the coefficients:
\beq \label{eq:relpp}
&\nonumber \strut\disp C_1=\alpha_1\sqrt {\frac {E_n+s_n}{2s_n}},\quad
C_2=-\alpha_2\frac {p_{z}}{\sqrt {2s_n(E_n+s_n)}}, & \\
&\nonumber \strut\disp C_1p_{z}+C_2b_n=\alpha_1(E_n+1)\frac {p_{z}}{\sqrt{2s_n(E_n+s_n)}}
,\quad C_1b_n-C_2p_{z}=\alpha_2(E_n+1)\sqrt {\frac {E_n+s_n}{2s_n}}. &
\eeq
From these relations we get $\alpha_1=(s_n+1)\alpha_0$, $\alpha_2= b_n\alpha_0$, where $\alpha_0$
have to be found from the normalization condition. Since the functions $\chi$ are normalized, one
can write down:
$$
\intl_{-\infty}^\infty[v_{n+}^{+}(p_{z},u)]^\dagger\gamma^0v_{n+}^{+}(p_{z},u)\d u=
\alpha_0^2[(s_n+1)^2-b_n^2]=\alpha_0^22(s_n+1)=1.
$$

2) For the electron with the helicity $-1$ we find the relations:
$$
C_1v^{+}_{n1}(p_{z},u)+C_2v^{+}_{n2}(p_{z},u)=v_{n-}^{+}(p_{z},u)=
\alpha_1U_2\chi_{n-1}+\alpha_2U_3\chi_n,
$$
and then the relations for the coefficients:
\beq \label{eq:relpm}
&\nonumber \strut\disp C_1=\alpha_1\frac {p_{z}}{\sqrt {(E_n+s_n)2s_n}},\quad
C_2=\alpha_2\sqrt {\frac {E_n+s_n}{2s_n}}, & \\
&\nonumber \strut\disp C_1p_{z}+C_2b_n=-\alpha_1(E_n+1)\sqrt {\frac {E_n+s_n}{2s_n}},
\quad C_1b_n-C_2p_{z}=\alpha_2(E_n+1)\frac {p_{z}}{\sqrt {2s_n(E_n+s_n)}}. &
\eeq
Then $\alpha_1=b_n\alpha_0,\,\,\alpha_2=(s_n+1)\alpha_0$ and $\alpha_0$
is the same as for the previous case since
$$
\intl_{-\infty}^\infty[v_{n-}^{+}(p_{z},u)]^\dagger\gamma^0v_{n-}^{+}(p_{z},u)\d u=
\alpha_0^2[b_n^2-(s_n+1)^2]=\alpha_0^22(s_n+1)=1.
$$

3) For the positron with the helicity $-1$:
$$
C_1v^{-}_{n1}(-p_{z},u)+C_2v^{-}_{n2}(-p_{z},u)=v_{n+}^{-}(p_{z},u)=
\alpha_1U_1\chi_{n-1}+\alpha_2U_4\chi_n.
$$
The relations for the coefficients: 
\beq \label{eq:relmp}
&\nonumber \strut\disp C_1=\alpha_1\frac {p_{z}}{\sqrt {(E_n+s_n)2s_n}},\quad
C_2=\alpha_2\sqrt {\frac {E_n+s_n}{2s_n}}, & \\
&\nonumber \strut\disp C_1p_{z}-C_2b_n=\alpha_1(E_n+1)\sqrt {\frac {E_n+s_n}{2s_n}},
\quad -C_1b_n-C_2Z=-\alpha_2(E_n+1)\frac {p_{z}}{\sqrt {2s_n(E_n+s_n)}} &
\eeq
and $\alpha_1=b_n\alpha_0,\,\,\alpha_2=-(s_n+1)\alpha_0$. 

4) For the pozitron with the helicity $+1$:
$$
C_1v^{-}_{n1}(-p_{z},u)+C_2v^{-}_{n2}(-p_{z},u)=v_{n-}^{-}(p_{z},u)=
\alpha_1U_3\chi_{n-1}+\alpha_2U_2\chi_n.
$$

As a result we get the expressions for the fixed helicity in a form which we would use in 
the final expressions for the solution of the Dirac equation:
\beq \label{eq:vnpp}
& \strut\disp v_{n+}^{+}(p_{z},u)=\frac {1}{\sqrt {2}}\left[\sqrt {s_n+1}
U_1(p_{z})\chi_{n-1}(u)+\sqrt {s_n-1}U_4(p_{z})\chi_n(u)\right], & \\
\label{eq:vnpm}
& \strut\disp v_{n-}^{+}(p_{z},u)=\frac {1}{\sqrt {2}}\left[\sqrt {s_n-1}
U_3(p_{z})\chi_{n-1}(u)+\sqrt{s_n+1}U_2(p_{z})\chi_n(u)\right], & \\ \label{eq:vnmp}
& \strut\disp v_{n+}^{-}(-p_{z},u)=\frac {1}{\sqrt {2}}\left[\sqrt {s_n+1}
U_3(-p_{z})\chi_{n-1}(u)-\sqrt{s_n-1}U_2(-p_{z})\chi_n(u)\right], & \\ \label{eq:vnmm}
& \strut\disp v_{n-}^{-}(-p_{z},u)=\frac {1}{\sqrt {2}}\left[-\sqrt {s_n-1}
U_1(-p_{z})\chi_{n-1}(u)+\sqrt{s_n+1}U_4(-p_{z})\chi_n(u)\right], &
\eeq
where $U_i(p)$ are defined by equations (\ref{eq:Ui}). The spinors (\ref{eq:vnpp}-\ref{eq:vnmm}) are used in equation (\ref{eq:Sxxif}) for calculation of the $S$-matrix elements.

\subsection{Particular and total solution for electron in a strong magnetic field}
\label{sec:Solution}

The particular solutions of eq. (\ref{eq:eqDPsi}) could be written in the following form:
\be \label{eq:Psisol}
\Psi_{n s}^{\eps}(\ur,p_{y},p_{z})=
\frac {1}{2\pi\sqrt {E_n(p_{z})}}\,v_{n s}^{\eps}(\eps p_{z},x+\eps p_{y}/b)
\exp {[-\eps\,i(E_n\,t-p_{y}\,y-p_{z}\,z)]},
\ee
where $v_{n s}^{\eps}(\eps p_{z},u)$ are defined by equations (\ref{eq:vnpp}-\ref{eq:vnmm}). This solution is used in construction of the $S$-matrix element (\ref{eq:Sfi}) and relativistic electron propagator (\ref{eq:prop}).
The spinors (\ref{eq:vnpp})--(\ref{eq:vnmm}) compose an orthonormal system and
$
\int_{-\infty}^\infty v_{n s}^{\eps\dagger}(\eps p_{z},u)
v_{n's'}^{\eps'}(\eps p_{z},u)\d u=
E_n(p_{z})\,\delta_{nn'}\delta_{\eps\,\eps'}\delta_{s s'}.
$
Therefore it is easy to find the relations of orthonormality for the solutions of the Dirac equation:
\be \label{eq:Psiort}
\int\d^3r\Psi_{n s}^{\eps\dagger}(\vr,t,p_{y},p_{z})
\Psi_{n's'}^{\eps'}(\vr,t,Y',p_{z}')=\delta(p_{y}-p_{y}')\delta(p_{z}-p_{z}')\delta_{nn'}
\delta_{\eps\eps'}\delta_{s s'}.
\ee
The condition of completeness of the system takes form
\be \label{eq:Psiful} 
\sum_{n,\eps,s}\int\d p_{y}\,\d p_{z}\Psi_{n s}^{\eps}(\vr,t,p_{y},p_{z})
\Psi_{n s}^{\eps\dagger}(\vr\,',t,p_{y},p_{z})=\delta(\vr-\vr\,').
\ee

Therefore we can get the solution of the Cauchy problem with the initial function $\Phi(\vr,t_0)$ 
as an expansion over the particular solutions: 
\be \label{eq:Caushi}
\Phi(\vr,t)=\sum_{n,\eps,s}\int\d p_{y}\,\d p_{z}
\Psi_{n s}^{\eps}(\vr,t-t_0,p_{y},p_{z})\int\d^3\,r'
\Psi_{n s}^{\eps\dagger}(\vr\,',t-t_0,p_{y},p_{z})\,\Phi(\vr\,',t_0).
\ee

%

These wave functions given by (\ref{eq:Psisol}) satisfy the equations
$$
\left[i\left(\gamma^0\Dr {}{t}+\gamma^1\Dr {}{x}+\gamma^2\Dr {}{y}+
\gamma^3\Dr {}{z}\right)-bx\gamma^2-1\right]\Psi_{n s}^\eps(\ur,p_{y},p_{z})=0,
$$
while the spinors are the solution of equations:
$$
\left(i\gamma^1\DR {}{u_\eps}-bu_\eps\gamma^2+\eps E_n(p_{z})\gamma^0-
\eps p_{z}\gamma^3-1\right)v_{n s}^\eps(\eps p_{z},u_\eps)=0.
$$

\section{Expressions for spinor products}
\label{sec:ESpPr}

Expression (\ref{eq:regprop}) contains only separate spinor products and
therefore there are more terms than in the non regularized case (\ref{eq:proectel})--(\ref{eq:proectps}). Nevertheless the analytical expressions for the products can be found:
\beq \label{eq:fracmat}
& \strut\disp v_{n+}^{+}(p_{z},x_2)v_{n+}^{+\dagger}(p_{z},x_1)=\frac {1}{4s_n}
\left[(s_n+1)\chi_{n-1}(x_2)\chi^*_{n-1}(x_1)(E_n\Sigma^{+}+p_{z}\alpha^{+}+
s_nD^{+})+ \right. & \nonumber \\
& \strut\disp (s_n-1)\chi_n(x_2)\chi^*_n(x_1)(E_n\Sigma^{-}-
p_{z}\alpha^{-}-s_nD^{-})+ & \nonumber \\
& \strut\disp \left. b_n\left(\chi_{n-1}(x_2)\chi^*_n(x_1)
(E_n\gamma_{+}-p_{z}D_{+}+s_n\alpha_{+})-\chi_n(x_2)\chi^*_{n-1}(x_1)(E_n\gamma_{-}
+p_{z}D_{-}-s_n\alpha_{-})\right)\right], & \\
& \strut\disp v_{n-}^{+}(p_{z},x_2)v_{n-}^{+\dagger}(p_{z},x_1)=\frac {1}{4s_n}
\left[(s_n-1)\chi_{n-1}(x_2)\chi^*_{n-1}(x_1)(E_n\Sigma^{+}+p_{z}\alpha^{+}-
s_nD^{+})+ \right. & \nonumber \\
& \strut\disp (s_n+1)\chi_n(x_2)\chi^*_n(x_1)(E_n\Sigma^{-}-
p_{z}\alpha^{-}+s_nD^{-})+ & \nonumber \\
& \strut\disp \left. b_n\left(\chi_{n-1}(x_2)\chi^*_n(x_1)
(-E_n\gamma_{+}+p_{z}D_{+}+s_n\alpha_{+})+\chi_n(x_2)\chi^*_{n-1}(x_1)
(E_n\gamma_{-}+p_{z}D_{-}+s_n\alpha_{-})\right)\right], & \\
& \strut\disp v_{n+}^{-}(p_{z},x_2)v_{n+}^{-\dagger}(p_{z},x_1)=\frac {1}{4s_n}
\left[(s_n-1)\chi_{n-1}(x_2)\chi^*_{n-1}(x_1)(E_n\Sigma^{+}+p_{z}\alpha^{+}+
s_nD^{+})+ \right. & \nonumber \\
& \strut\disp (s_n+1)\chi_n(x_2)\chi^*_n(x_1)(E_n\Sigma^{-}-
p_{z}\alpha^{-}-s_nD^{-})- & \nonumber \\
& \strut\disp \left. b_n\left(\chi_{n-1}(x_2)\chi^*_n(x_1)
(E_n\gamma_{+}-p_{z}D_{+}+s_n\alpha_{+})+\chi_n(x_2)\chi^*_{n-1}(x_1)(E_n\gamma_{-}
+p_{z}D_{-}-s_n\alpha_{-})\right)\right], & \\
& \strut\disp v_{n-}^{-}(p_{z},x_2)v_{n-}^{-\dagger}(p_{z},x_1)=\frac {1}{4s_n}
\left[(s_n+1)\chi_{n-1}(x_2)\chi^*_{n-1}(x_1)(E_n\Sigma^{+}+p_{z}\alpha^{+}-
s_nD^{+})+ \right. & \nonumber \\
& \strut\disp (s_n-1)\chi_n(x_2)\chi^*_n(x_1)(E_n\Sigma^{-}-
p_{z}\alpha^{-}+s_nD^{-})+ & \nonumber \\
& \strut\disp \left. b_n\left(\chi_{n-1}(x_2)\chi^*_n(x_1)
(E_n\gamma_{+}-p_{z}D_{+}-s_n\alpha_{+})-\chi_n(x_2)\chi^*_{n-1}(x_1)
(E_n\gamma_{-}+p_{z}D_{-}+s_n\alpha_{-})\right)\right], &
\eeq
where the necessary designations are given in Appendix \ref{sec:DirMatr}.
These expressions are valid for both Feynman diagrams, but one should differentiate the values $V,\,E_n,\,p_{z},\,x_1,
\,x_2$ and $\Gamma_n$ for each of them according to the specific arguments in the expression for the $S$-matrix elements (\ref{eq:Sxxif}).

\end{document}